\documentclass[12pt,twoside,aps,showkeys,showpacs]{revtex4-1}
\usepackage[english]{babel}
\usepackage{epsf,epsfig,graphics}
\usepackage{amssymb}
\usepackage{amsmath}
\usepackage{amsthm}
\usepackage{latexsym}
\def \slash#1{\not\! #1}
\begin{document}

\title{Relativistic potential of a hydrogen-like system in Poincar\'{e} invariant quantum mechanics
}

\author{V.V. Andreev}
\affiliation{Francisk Skorina Gomel State University, Gomel, Belarus}, \\
\email{vik.vandreev59@gmail.com}           

\begin{abstract}
\hspace{5mm}

To describe a relativistic hydrogen atom we used the Poincar\'{e}-covariant model
of a two particle system with gauge invariant potential. The kernel of the radial
integral equation is obtained which describes a system of two fermions with
electromagnetic interaction.
\end{abstract}


\maketitle

\section{Introduction}

The investigation of the energy spectra of hydrogenic atoms is of great importance
for high accuracy verification of the Standard Model and derivation of more correct
values of fundamental physical constants (the fine structure constant, the masses
of the muon and electron, the proton charge radius, etc.)
\cite{Eides:2000xc,Karshenboim:2003vs,Karshenboim:2006ht}.

The experiment to measure the energy interval $E(2P_{3/2}^{F=2})- E(2S_{1/2}^{F=
1})$ of the muonic hydrogen atom \cite{Pohl2010er} led to a significant difference
with theoretical calculations. This circumstance stimulated a new round of research
on the simplest atomic systems and their parameters (see, for example,
\cite{DeRujula:2010dp,Jentschura2011500,Jentschura2011516,Dorokhov:2020ubu}).

The important stimulus for these evaluation is provided  by the spectacular
experimental progress in measurements of two fermion system energy levels. The
relative uncertainty of the atomic energy levels' frequency measurement was
reduced to $4.5 \times 10^{-15}$ \cite{PhysRevLett.110.230801}.

The basis for calculating the energy structure of coupled systems is the procedure
for constructing the relativistic potential of particle interaction. The
construction of the operator for the interaction potential of a particles system
is carried out, as a rule, with the help of the corresponding amplitude $T_{\rm f\/
i}$ of elastic scattering \cite{Akhiezer:1965engl,Pilkuhn1979,Lucha:1991jy}.

The most common technique is to calculate spinor structures in terms of Pauli
matrices and momenta using explicit bispinors
\cite{Lucha:1991jy,Galkin:1992ry,Crater:1996ti,Terekidi:2003gp}. Such a
calculation, as a rule, is done approximately, using the expansion in the
velocities $\upsilon /c $ of the particles of the system. Next, the potential $
V({\bf r})$ in the coordinate space is calculated as the Fourier transformation of the
above scattering amplitude $T_{\rm f\/ i}$.

In contrast to the above-mentioned technique, the work plans to make an accurate
calculation (without the expansion in the velocities of the system particles) of
the kernel of the radial equation of the relativistic system as a scalar function
of the momenta of the particles. In this situation, it is advisable to use methods
of direct calculation of the corresponding matrix elements as explicitly scalar
functions.

The aim of this work is to calculate the kernel of the radial equation of a
two-fermion relativistic system in momentum space based on an accurate calculation
of the amplitude of one-boson exchange.

This calculation is not straightforward and therefore represents an independent
task. Calculation of such a kernel will be possible in the future to refine
the contributions of higher-order relativistic effects to the energy spectra of
hydrogen-like systems.

\section{Two-particle system in Poincar\'{e} invariant quantum mechanics}

For calculation of fermion-fermion systems energy levels with electromagnetic
interaction there exist a number of various models: Bete-Salpeter equation \cite{Bete:1951,Salpeter:1952}, model
based on effective Dirac equation \cite{Eides:2000xc}, quasipotential approach
\cite{Faustov:1998kx,Faustov:1997rc,Martynenko:2006gz}, variational Hamiltonian
formalism \cite{Terekidi:2003gp} and others.

In our article we use the description of bound states with the help of  the
Poincar\'{e}-invariant quantum mechanics \cite{Polyzou:2010kx} (or relativistic
Hamiltonian dynamics (RHD) \cite{Keister:1991sb}). In this approach, the
Hamiltonian $\hat{H}$ is assumed to be the sum of a relativistic kinetic energy
operator $T\left(\mathbf{k}\right)$ that represents the invariant mass of two
noninteracting particles plus phenomenological interaction $\hat{V}$. The kinetic
energy operator has the form
\begin{equation}
T\left(\mathbf{k}\right) \equiv M_0 = \sqrt{m_1^2+{\bf k}^2} +
\sqrt{m_2^2+{\bf k}^2} ~, \label{va1}
\end{equation}
where $ \mathbf{k}$ is the relative momentum. The total momentum of the free-system
$\mathbf{P}$ is
\begin{equation}\label{va3}
\mathbf{P}=\mathbf{p_1+p_2}\; ,
\end{equation}
and $\omega_m\left(\mathrm{p}\right)=\sqrt{m^2+{\bf p}^2}$,
$\mathrm{k}=\left|\mathbf{k}\right|$.

In this approach the bound system with the momentum $\mathbf{Q}$, eigenvalues $E$ and
spin $J$ is described by the wave function $ \Phi ^{J \mu}_{\mathbf{Q};\;\sigma_{1}
\sigma_{2}}\left({\bf k}\right)$ of the two-particle state which satisfies the equation
\begin{eqnarray}
&& \sum_{\lambda_{1},\lambda_{2} }\int < {\bf k},\sigma_{1},\sigma_{2}\parallel\hat{
V}\parallel \ {\bf k}^{\prime}, \lambda_{1},\lambda_{2}>\Phi ^{J
\mu}_{\mathbf{Q};\lambda_{1} \lambda_{2}}\left({\bf k}^{\prime }\right)\mathrm{d} {\bf
k}^{\prime}=
\nonumber\\
&& =\left(E-M_0\right) \Phi ^{J}_{\mathbf{Q};\sigma_{1} \sigma_{2}}\left({\bf
k}\right) \;.\label{maineq}
\end{eqnarray}
The radial equation for the two-particle bound state in the center-momen\-tum system has
the following form
\begin{equation}
\sum_{\ell^{\prime},S^{\prime}} \int\limits_{0}^{\infty} V^{J}_{\ell,S\;
;\ell^{\prime}, S^{\prime}}\left(\mathrm{k},\mathrm{k^{\prime}}\right) \Phi^{J
\mu}_{\ell^{\prime}, S^{\prime}}\left(\mathrm{{k}^{\prime
}}\right){\mathrm{{k}^{\prime}}}^{2}\mathrm{d}{\mathrm{k}}^{\prime}= \left(M-
M_0\right) \Phi ^{J}_{\ell, S}\left(\mathrm{{k}}\right)\;. \label{maineq1}
\end{equation}
The equation (\ref{maineq1}) can be obtained with the help of the Clebsh-Gordan
coefficients of Poincar\'{e} group (see,\cite{Keister:1991sb}). The corresponding
potential operator $V^{J}_{\ell^{\prime}, S^{\prime},\ell,
S}\left(\mathrm{k^{\prime}},\mathrm{k}\right)=< \mathrm{{k}^{\prime}},
J,\mu,\ell^{\prime}, S^{\prime}
\parallel \hat{V}\parallel  \mathrm{{k}}, J, \mu, \ell, S >$
is obtained from the equation
\begin{eqnarray}
&&V^{J}_{\ell^{\prime},S^{\prime}\;
;\ell,S}\left(\mathrm{k^{\prime}},\mathrm{{k}}\right)=\frac{\sqrt{\left(2\ell+1\right)
\left(2\ell^{\prime}+1\right) }}{2J+1} \sum_{\lambda_{1}, \lambda_{2},
\lambda_{1}^{\prime}, \lambda_{2}^{\prime}} {\mathbf{C} \left\{ \smallmatrix
\hspace{-2pt} 1/2 &\hspace{-2pt} 1/2 &\hspace{-2pt} S \\
\hspace{-2pt} \lambda_{1},&\hspace{-2pt} -\lambda_{2},&\hspace{-2pt} \lambda
\endsmallmatrix
\right\}} {\mathbf{C} \left\{ \smallmatrix
\hspace{-2pt}{\ell} &\hspace{-2pt} S &\hspace{-2pt} J \\
\hspace{-2pt} 0 ,&\hspace{-2pt}\lambda, &\hspace{-2pt} \lambda
\endsmallmatrix
\right\}}\times\nonumber\\
&& \times   {\mathbf{C} \left\{ \smallmatrix
\hspace{-2pt}{\ell^{\prime}} &\hspace{-2pt} S^{\prime} &\hspace{-2pt} J^{\prime} \\
\hspace{-2pt} 0 ,&\hspace{-2pt}\lambda^{\prime},&\hspace{-2pt} \lambda^{\prime}
\endsmallmatrix
\right\}} {\mathbf{C} \left\{ \smallmatrix
\hspace{-2pt} 1/2 &\hspace{-2pt} 1/2 &\hspace{-2pt} S \\
\hspace{-2pt} \lambda_{1}^{\prime},&\hspace{-2pt}
-\lambda_{2}^{\prime},&\hspace{-2pt} \lambda
\endsmallmatrix
\right\}} < \mathrm{k}^{\prime},J,\mu,\lambda_{1}^{\prime}, \lambda_{2}^{\prime}
\parallel \hat{V}\parallel  \mathrm{k}, J, \mu,\lambda_{1},\lambda_{2}>\;,
\label{redmat1}
\end{eqnarray}
The matrix element
$V^{J,J^{\prime}}_{\lambda_{1}^{\prime},\lambda_{2}^{\prime},\lambda_{1},\lambda_{2}}
\left(\mathrm{k^{\prime}},\mathrm{k}\right)= <
\mathrm{k}^{\prime},J^{\prime},\mu^{\prime},\lambda_{1}^{\prime},\lambda_{2}^{\prime}
\parallel \hat{V}\parallel  \mathrm{k}, J, \mu,\lambda_{1},\lambda_{2}>$ is related
with $< {\bf k}^{\prime},\lambda_{1}^{\prime},\lambda_{2}^{\prime}\parallel \hat{V}
\parallel {{\bf k}}, \lambda_{1},\lambda_{2}>$ by the means of the Jacob-Wick
decomposition (see, for example, \cite{Brown79engl}) and has the form
\begin{eqnarray}
&&V^{J,J^{\prime}}_{\lambda_{1}^{\prime},\lambda_{2}^{\prime},\lambda_{1},\lambda_{2}}
\left(\mathrm{k^{\prime}},\mathrm{k}\right)=
\frac{\sqrt{\left(2J+1\right)\left(2J^{\prime}+1\right)
}}{4\pi} \times \nonumber\\
&& \times \int \mathrm{d}^{2}\hat{\mathbf{k}}\;
\mathrm{d}^{2}\mathbf{\hat{k}}^{\prime}\;D_{\mu^{\prime}\;\lambda^{\prime}}^{J^{\prime}}
\left(\varphi_{k^{\prime}},\theta_{k^{\prime}},-\varphi_{k^{\prime}}\right)
D_{\mu\;\lambda}^{\ast\;J} \left(\varphi_{k},\theta_{k},-\varphi_{k}\right) \times \nonumber\\
&& \times \left\langle \mathbf{k}^{\prime}, \lambda _{1}^{\prime},
\lambda _{2}^{\prime} \right.
\parallel \hat{V} \left. \parallel \mathbf{k}, \lambda _{1},\lambda _{2} \right\rangle
\;, \label{hel1}
\end{eqnarray}
with $\lambda=\left(\lambda_{1}-\lambda_{2}\right)/2 $. Functions ${\mathbf{C}
\left\{ \smallmatrix
\hspace{-2pt} s_1 &\hspace{-2pt} s_2 &\hspace{-2pt} S \\
\hspace{-2pt} \lambda_{1},&\hspace{-2pt} \lambda_{2},&\hspace{-2pt} \lambda
\endsmallmatrix
\right\}}$, are Clebsh-Gordan coefficients of the $SU(2)$-group and the function
$D_{\mu\;\lambda}^{J}\left(\varphi_{k},\theta_{k},-\varphi_{k}\right)$ is the Wigner
$D$-function with the angle of vector
$\mathbf{\hat{k}}=\mathbf{k}/\left|\mathbf{k}\right|$.

Using the Wigner-Eckart theorem we have the Eq.(\ref{hel1}) transformed into
\begin{eqnarray}
&&
V^{J,J^{\prime}}_{\lambda_{1}^{\prime},\lambda_{2}^{\prime},\lambda_{1},\lambda_{2}}
\left(\mathrm{k^{\prime}},\mathrm{k}\right)= \delta_{J,J^{\prime}}
\;\delta_{\mu,\mu^{\prime}}\; V^{J}_{\lambda_{1}^{\prime},
\lambda_{2}^{\prime}\;;\lambda_{1},\lambda_{2}}\left(\mathrm{k}^{\prime},\mathrm{k}\right)\;,
\nonumber\\
&&  V^{J}_{\lambda_{1}^{\prime},
\lambda_{2}^{\prime};\lambda_{1},\lambda_{2}}\left(\mathrm{k}^{\prime},\mathrm{k}\right)=
\nonumber\\
&& =\int\limits_{-1}^{1}\mathrm{d}\left(\cos \beta\right)\int\limits_{0}^{2\pi}
\;\mathrm{d}\varphi\;D_{\lambda,\lambda^{\prime}}^{J} \left(\varphi,\beta,-\varphi
\right) \left\langle \mathbf{k}^{\prime}, \lambda _{1}^{\prime}, \lambda _{2}^{\prime}
\right.
\parallel \hat{V} \left. \parallel \mathbf{k}, \lambda _{1},\lambda _{2}
\right\rangle \;, \label{hel2}
\end{eqnarray}
where
\begin{equation}
\cos\beta = \left(\mathbf{k}\mathbf{k}^{\prime}\right)/
\left(\left|\mathbf{k}\right|\left|\mathbf{k}^{\prime}\right|\right)= \cos
\theta_{k^{\prime}}\cos \theta_{k}+\cos \left(\varphi_{k^{\prime}}-\varphi_{k} \right)
\sin \theta_{k^{\prime}}\sin \theta_{k} \;.\label{va18}
\end{equation}

\section{Relativistic potential of hydrogen-like systems}
The interaction potential is constructed with the help of the scattering amplitude
according to the below prescription \cite{Lucha:1991jy}:
\begin{description}
  \item[1.] Compute the scattering amplitude $R_{f i}$, which is
  defined in terms of the $S$-matrix element by the decomposition
\begin{equation}\label{va10}
S_{f i}=\delta_{f i}-\mathrm{i}\left(2\pi\right) \delta\left(E_f-E_i\right)R_{f i},
\end{equation}
where $i$ and $f$ denote initial and final states respectively.
 \item[2.] The potential $\hat{V}$ can be extracted with the help of the relation
\begin{equation}\label{va11}
R_{f i}=\left\langle f \right|\hat{V}\left|i\right\rangle+\sum_{n
\neq i}\frac{\left\langle f \right|\hat{V}\left|n\right\rangle
R_{n i}}{E_{i}-E_{n}+\mathrm{i} \epsilon}\;.
\end{equation}
or the operator form
\begin{equation}\label{va11as}
\hat{V}=\frac{R}{\mathrm{I}+G_{red}R}\;,
\end{equation}
where $G_{red}$ is the reduced Green function
\begin{equation}\label{va11bs}
G_{red}=\sum_{n \neq i}^{\infty}\frac{\left|n\right\rangle
\left\langle n \right|}{E_{i}-E_{n}+\mathrm{i} \epsilon}\;.
\end{equation}
\end{description}

Therefore, by investigating the corresponding scattering problem of bound-state
constituents the potential (or a part of the potential) may be derived according to the
above-mentioned procedure. Let us illustrate the recipe of a potential calculation by
applying it to the electron-proton bound system (Hydrogen atom).

We start with the process
\begin{equation}\label{va12}
e^-(k_1,\lambda_{k_1}) + p(k_2,\lambda_{k_2}) \rightarrow e^-(p_1,\lambda_{p_1}) +
p(p_2,\lambda_{p_2}) \;,
\end{equation}
where the momenta of the particles and spin numbers $\left( \lambda_{k_i}= \pm 1,
\lambda_{p_i}= \pm 1 \right)$ are given between the parentheses.

According to the perturbation theory, the matrix element of the
fermion-fermion system potential will be represented by a series of matrix elements with
respect to the fine structure constant $\alpha$, where the main contribution
determines the one-photon exchange between $e^-$ and $p$.

\begin{figure}[h t b]
\centering  \resizebox{0.7\textwidth}{!}{
\includegraphics{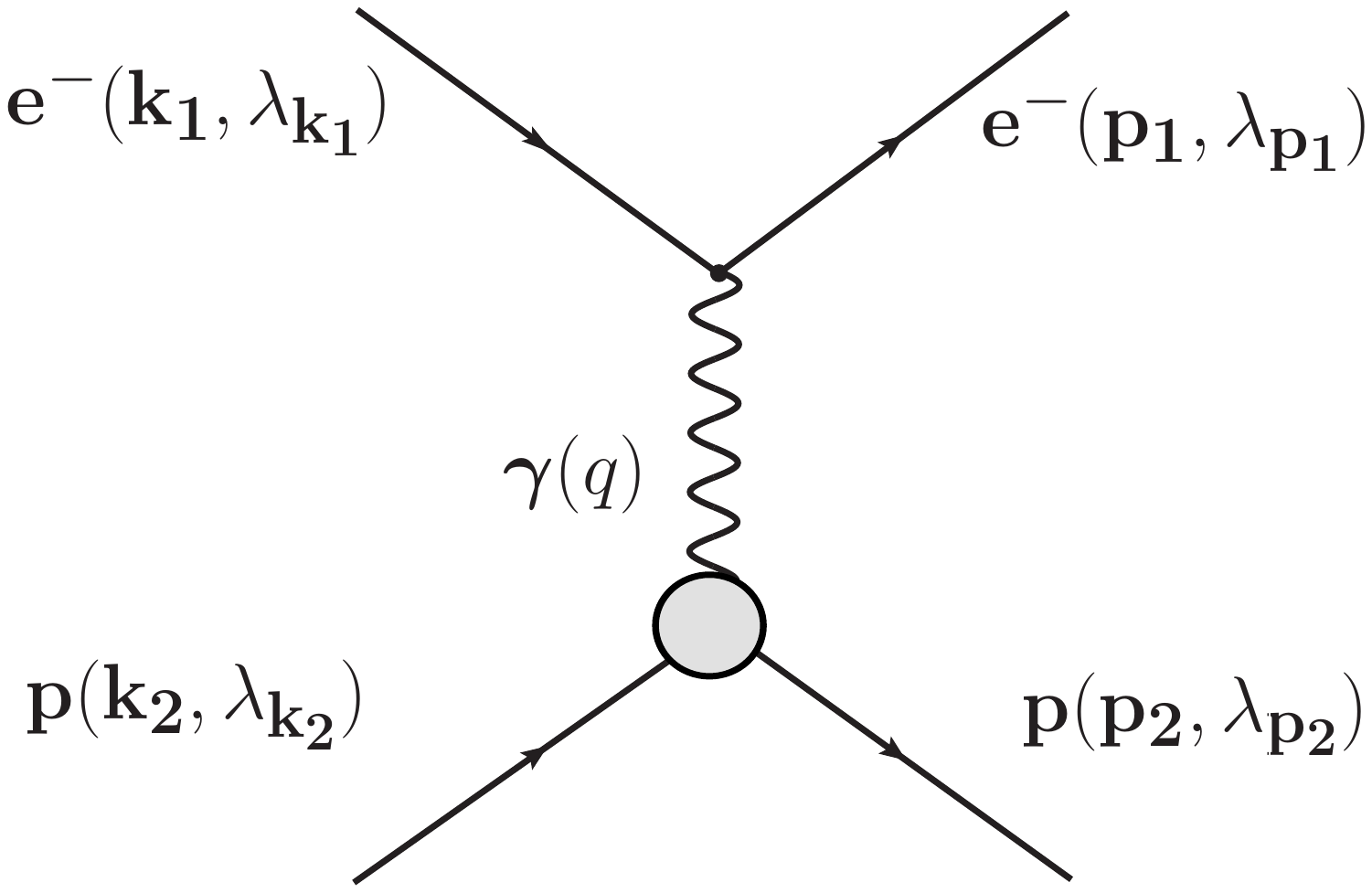}}
\vspace{-2mm} \caption{One-photon exchange Feynman diagram of $e^-\;p$ elastic
scattering} \label{fig1}
\end{figure}

The initial approximation of the potential $V$ for a bound system was selected in the
form of the potential which corresponds to the tree-level diagram, as depicted in
Fig.~\ref{fig1}

Using the Feynman rules we write down the matrix element corresponding to the
one-photon exchange diagram \ref{fig1}:
\begin{eqnarray}
&& <{\bf k}^{\prime},\lambda_{p_1},\lambda_{p_2}\parallel \hat{ V}
\parallel \ {\bf k}, \lambda_{k_1},\lambda_{k_2}>=V_{1\gamma}(\mathbf{k},
\mathbf{k}^{\prime})=\nonumber\\&& =(-1)\;N_{\mathrm{k},\mathrm{k^{\prime}}}\frac{Z
\;\alpha}{8 \pi^{2}\;q^2}\; j_{\lambda
_{p_1},\lambda_{k_1}}^{\mu}\left(p_1,k_1\right) D_{\mu \nu}\left(q\right)j_{\lambda
_{p_2},\lambda_{k_2}}^{\nu}\left(p_2,k_2\right) \;, \label{onephoton}
\end{eqnarray}
where
\begin{eqnarray}
&&\hspace{-15mm} j_{\lambda _{p_i},\lambda_{k_i}}^{\mu}\left(p_i,k_i\right) =
\bar{u}_{\lambda _{p_i}}\left(p_{i}\right)\gamma^{\mu}
u_{\lambda_{k_i}}\left(k_i\right)\;\;\left(i=1,2\right)\;. \label{toki}
\end{eqnarray}

Four-momenta of particles in the center of mass frame have components
\begin{eqnarray}
&&k_1=\left(\omega _{m_1}\left(\mathrm{k} \right),\;\;\;\mathbf{k}\right)\;,
p_1=\left(\omega _{m_1}\left(\mathrm{k}^{\prime } \right),\;\;\;\mathbf{k}^{\prime
}\right),\;
\nonumber\\
&&k_2=\left(\omega _{m_2}\left(\mathrm{k} \right),\;-\mathbf{k}\right),\;
p_2=\left(\omega _{m_2}\left(\mathrm{k}^{\prime }\right),\;-\mathbf{k}^{\prime
}\right)\;.\label{va14}
\end{eqnarray}
The parameter $Z$ specifies the value of the electric charge of the second fermion
(for a hydrogen atom $Z$ = 1), while
$$
N_{\mathrm{k},\mathrm{k^{\prime}}}=1/\sqrt{\omega_{m_1}\left(\mathrm{k} \right)
\omega_{m_1}\left(\mathrm{k}^{\prime}\right) \omega_{m_2}\left(\mathrm{k}\right)
\omega_{m_2}\left(\mathrm{k}^{\prime}\right)}\;.
$$

The function $D_{\mu \nu}\left(q\right) $ is associated with the photon propagator
(without $q^2$ and normalization factors)
\begin{equation}
\label{propphoton}
D_{\mu\nu}\left(q\right)=\left(g_{\mu\nu}-\frac{q_{\mu}q_{\nu}}{q^2}
+\xi_{A}\;\frac{q_{\mu}q_{\nu}}{q^2}\right)\;,
\end{equation}
where $\xi_{A}$ is the gauge parameter and $q$ is the photon momentum in the
center of mass system:
\begin{equation}\label{qfot}
q=\left\{q_0=0, \; \;  \mathbf{k}-\mathbf{k}^{\prime}\right\}\;.
\end{equation}

For the potential with $V_{1\gamma}(\mathbf{k}, \mathbf{k}^{\ prime})$ it must be modified
in order to satisfy the gauge invariance condition. The current conservation
requirement (see \cite{Pilkuhn1979}) leads to the fact that
\begin{equation}\label{cal1}
q_{\mu}\;j_{\lambda
_{p_1},\lambda_{k_1}}^{\mu}\left(p_1,k_1\right)=q_{\mu}j_{\lambda
_{p_2},\lambda_{k_2}}^{\mu}\left(p_2,k_2\right)=0\;.
\end{equation}
However, as follows from the definition of (\ref{toki}), the relations
\begin{eqnarray}
&&\hspace{-15mm}q _{1,\;\mu}\;j_{\lambda
_{p_1},\lambda_{k_1}}^{\mu}\left(p_1,k_1\right)=0\; \; ,\; \; \; \;\;
q_{2,\;\mu}\;j_{\lambda
_{p_2},\lambda_{k_2}}^{\mu}\left(p_2,k_2\right)=0\;,\;\mbox{with} \label{cal2}
\end{eqnarray}
\begin{eqnarray}
&&\hspace{-20mm} q_1 =\left\{\omega
_{m_1}\left(\mathrm{k}\right)-\omega_{m_1}\left(\mathrm{k}^{\prime}\right),
\;\mathbf{k}-\mathbf{k}^{\prime} \right\} , \; q_2 =\left\{\omega
_{m_2}\left(\mathrm{k}^{\prime}\right)-\omega
_{m_2}\left(\mathrm{k}\right),\;\mathbf{k}-\mathbf{k}^{\prime}\right\},
\label{q1q2}
\end{eqnarray}
are satisfied, but not the requirement (\ref{cal1}).

The reason for this is the difference between the components $q_0 = 0$, ${q_1}_0$
and ${q_2}_0$. This difference is due to the need to take into account the effects
of the virtuality of particles of the linked system \cite{Pilkuhn1979}. Therefore,
the relation $\mathrm {k}=\mathrm {k}^{\ prime}$ which follows from the
conservation laws of the 4-momentum  in the case of elastic scattering cannot be
applied to eliminate the mismatch of the zero components of the 4-vectors $q$ and $
q_1, q_2$.

The requirement of gauge invariance (\ref{cal1}) can be met by modifying the
currents $j_{\lambda_{p_{1,2}}, \lambda_{k_{1,2}}}^{\mu} \left(p_{1,2}, k_{1,2}
\right)$ by overriding them (see, for example, \cite{Klink:2003ec}):
\begin{equation}\label{newtok}
j_{\lambda _{p_{i}},\lambda_{k_{i}}}^{\mu}\left(p_{i},k_{i}\right) \rightarrow
j_{\lambda _{p_{i}},\lambda_{k_{i}}}^{\mu}\left(i\right)= \left(g^{\mu}_{\;
\nu}-\frac{q^{\mu}q_{\nu}}{q^2}\right) \;j_{\lambda
_{p_{i}},\lambda_{k_{i}}}^{\nu}\left(p_{i},k_{i}\right)\;.
\end{equation}

As a result, the potential of one-boson exchange for the Feynman gauge of photon,
taking into account the conservation of currents, will be written in the form
\begin{eqnarray}
&&V_{1\gamma}(\mathbf{k}, \mathbf{k}^{\prime})
=N_{\mathrm{k},\mathrm{k^{\prime}}}\frac{Z \;\alpha}{8 \pi^{2}\;q^2}\; j_{\lambda
_{p_1},\lambda_{k_1}}^{\mu}\left(1\right) \left(g^{\mu}_{\;
\nu}-\frac{q^{\mu}q_{\nu}}{q^2}\right) j_{\lambda _{p_2},\lambda_{k_2}}^{\nu}\left(
2\right)\;.  \label{onephotona}
\end{eqnarray}

Taking into account the contributions of higher-order diagrams in the interaction
constant $\alpha$ (vacuum polarization and photon exchange between electrons) we then
obtain the following expression for the potential
\begin{eqnarray}
&& <{\bf k}^{\prime},\lambda_{p_1},\lambda_{p_2}\parallel \hat{ V}
\parallel \ {\bf k}, \lambda_{k_1},\lambda_{k_2}>=-N_{\mathrm{k},\mathrm{k^{\prime}}}\frac{Z \alpha \;\Pi\left(
\alpha,q^2\right)}{8 \pi^{2}\hskip 1pt  q^2}\times\nonumber\\
&&\times \;J_{\lambda _{p_1},\lambda_{k_1}}^{\mu}\left(p_1,k_1\right) \left(g_{\mu
\hskip 1pt \rho}-\frac{q_{\mu}q_{\rho}}{q^2}\right)J_{\lambda
_{p_2},\lambda_{k_2}}^{\rho}\left(p_2,k_2\right) \;, \label{va13a}
\end{eqnarray}
where single-particle fermion currents are written in the form:
\begin{eqnarray}
&&J_{\lambda _{p_1},\lambda_{k_1}}^{\mu}\left(p_1,k_1\right) = \bar{u}_{\lambda
_{p_1}}\left(p_{1}\right) \left(F_{1}^{e}\left( q^2_{1}\right)
\gamma^{\mu}+\frac{F_{2}^{e}\left(q^2_{1} \right)}{2\;m_1}\;\mathrm{i}\hskip 1pt
\sigma^{\mu\hskip 1pt \nu}q_{1,\hskip 1pt \nu}\right) u_{\lambda
_{k_1}}\left(k_{1}\right)\;,\label{toks12a} \\
&& J_{\lambda _{p_2},\lambda_{k_2}}^{\mu}\left(p_2,k_2\right)=\bar{u}_{\lambda
_{p_2}}\left(p_{2}\right)\left(F_{1}^{p}\left(q^2_{2}\right) \gamma^{\mu}+
\frac{F_{2}^{p}\left(q^2_{2}\right)}{2\hskip 1pt m_2} \mathrm{i} \hskip 1pt
\sigma^{\mu\hskip 1pt \tau}q_{2, \tau}\right) u_{\lambda
_{k_2}}\left(k_{2}\right)\;. \label{toks12}
\end{eqnarray}

In equations (\ref{toks12a})  and (\ref{toks12}), the function $\Pi\left(\alpha,
q^2 \right)$ determines the contribution from the vacuum polarization due to
fermionic loops, and $F_{1,2}^{e, p} \left(q^2 \right) $ are the form factors of
the electron and proton. Explicit forms $\Pi\left(\alpha, q^2 \right)$ and
$F_{1,2}^{e}\left(q^2\right)$ can be found in e.g. \cite{Akhiezer:1965engl,Bjorken64}.

Proton electromagnetic form factors $F_{1,2}^{p}\left(q^2\right)$ are related to
Sachs form factors by means of the equation
\begin{eqnarray}
&&\hspace{-15mm}G_{E}^{p}\left(q^2\right)= F_{1}^{p}\left(q^2\right)+
\frac{q^{2}}{4 m_{p}^{2}} F_{2}^{p} \left(q^2\right)
\;,\;\;G_{M}^{p}\left(q^2\right)= F_{1}^{p}\left(q^2\right)+
F_{2}^{p}\left(q^2\right)\;. \label{saksformfactor}
\end{eqnarray}

The square of transfer momentum $q_{1,2}^2$
\begin{equation}\label{q2}
q_{1}^2=\left(p_1-k_1\right)^2 \;,\; \; \;  q_{2}^2=\left(k_2-p_2\right)^2\;
\end{equation}
can be represented by
\begin{equation}
\label{qz1} q^2_{1,2}=\left(1+\delta q_{1,2}^2\right) q^2\;,\;\;\mbox{where}
\end{equation}
\begin{eqnarray}
&& \left|\delta q_{1,2}^{2}\right|=\left|\frac{\left(\omega _{m_{1,2}}\left(
\mathrm{k} \right)-\omega
_{m_{1,2}}\left(\mathrm{k}^{\prime}\right)\right)^{2}}{q^2}\right| < 1
\;.\label{dd12}
\end{eqnarray}

Hence the contributions $\sim \delta q_{1,2}^{2}$ can be considered as corrections
for the leading contribution (see, \cite{Bodwin:1987mj}) where
\begin{equation}\label{pribl1}
q_{1,2}^2= q^2\;.
\end{equation}

\section{Calculation of the spinor part of the potential}

To simplify the calculations, we transform the currents (\ref{toks12}) using the
Gordon identity. Then, the first part of the potential (\ref{va13a}) that is proportional
to the scalar product of fermion currents is reduced to the sum:
\begin{eqnarray}
&& V_{\lambda_{p_1},\lambda_{p_2}\;
\lambda_{k_1},\lambda_{k_2}}^{\left(\mathrm{A}\right)}\left({\bf k}^{\prime}, {\bf
k} \right) =-\;\frac{Z \alpha}{8 \pi^{2}
q^2}\;N_{\mathrm{k},\mathrm{k^{\prime}}}\left. \bigg \{\right.
K^{\mathrm{\left(I\right) }}\left(\tilde{q}^2\right)\times
\nonumber\\
&&\times \bar{u}_{\lambda_{p_1}}\left(p_{1}\right) \gamma_{\mu}u_{\lambda
_{k_1}}\left(k_{1}\right)\bar{u}_{\lambda _{p_2}}\left(p_{2}\right)\gamma^{\mu}
u_{\lambda _{k_2}}\left(k_{2}\right)+ \frac{K^{\mathrm{\left(IV\right)
}}\left(\tilde{q}^2\right)}{4\;m_1 m_{2}}\times
\nonumber\\
&&
\times\left(p_{1}+k_{1}\right)^{\nu}\left(p_{2}+k_{2}\right)_{\nu}\bar{u}_{\lambda
_{p_1}}\left(p_{1}\right) u_{\lambda _{k_1}}\left(k_{1}\right) \bar{u}_{\lambda
_{p_2}}\left(p_{2}\right) u_{\lambda _{k_2}}\left(k_{2}\right)-
\nonumber\\
&&- \frac{K^{\mathrm{\left(II\right)
}}\left(\tilde{q}^2\right)}{2\;m_1}\bar{u}_{\lambda _{p_1}}\left(p_{1}\right)
u_{\lambda _{k_1}}\left(k_{1}\right)\bar{u}_{\lambda
_{p_2}}\left(p_{2}\right)\left(\hat{p}_1+\hat{k}_1\right) u_{\lambda
_{k_2}}\left(k_{2}\right)-
\nonumber\\
&&- \frac{K^{\mathrm{\left(III\right)
}}\left(\tilde{q}^2\right)}{2\;m_{2}}\bar{u}_{\lambda
_{p_1}}\left(p_{1}\right)\left(\hat{p}_2+\hat{k}_2\right) u_{\lambda
_{k_1}}\left(k_{1}\right)\bar{u}_{\lambda _{p_2}}\left(p_{2}\right) u_{\lambda
_{k_2}}\left(k_{2}\right)\left.\right.\bigg \}\; \label{vector1}
\end{eqnarray}
with functions depending on $q^2=-(\mathbf{k}-\mathbf{k}^{\prime})^{2}$ and
$\mathrm{k},~ \mathrm{k}^{\prime}$:
\begin{equation}
\label{funk1} K^{\mathrm{\left(I,II \right)
}}\left(\tilde{q}^2\right)=\Pi\left(\alpha,q^2\right)
G_{M}^{p}\left(q^2_{2}\right)\;\left\{ G_{M}^{e}\left(q^2_{1}\right)\;,
F_{2}^{e}\left(q^2_{1}\right) \right \} \;,
\end{equation}
\begin{equation}
\label{funk4} K^{\mathrm{\left(III,IV \right)
}}\left(\tilde{q}^2\right)=\Pi\left(\alpha,q^2\right)
\;F_{2}^{p}\left(q^2_{2}\right)\left\{G_{M}^{e}\left(
q^2_{1}\right),F_{2}^{e}\left(q^2_{1}\right)\right\}\;,
\end{equation}
and the magnetic form factor of fermions $G_{M}^{e,p}\left(q^2\right)$.

To convert spinor structures into explicitly scalar functions we use the method of
basis spinors (\texttt{MBS}) \cite{Andreev2003ce,Andreev:2004bn}. A detailed
description of the \texttt{MBS} elements can be found in the appendix \ref{app1}

With its help fermionic ``chains'' with the operator $\gamma^{\mu}$ are
represented as:
\begin{eqnarray}
&& \bar{u}_{\lambda _p}\left( p,s_p\right)\gamma^{\mu}\;u_{\lambda
_k}\left(k,s_k\right)= \sum_{\sigma,\rho = -1}^1 \sum_{A,C=-1}^1
\bar{s}^{\left(C\right) }_{\lambda_p, \sigma}\left(p\right) \Gamma^{C,A}_{\sigma
,\rho }\left[\gamma^{\mu}\right]s_{\rho, \lambda_k}^{\left(A\right)}
\left(k\right)\;, \label{acat2}
\end{eqnarray}
where the expansion coefficients in basis spinors $s_{\rho,
\lambda_k}^{\left(A\right)} \left(k\right)$ for helicity states are defined by the
relations
\begin{eqnarray}
&& s^{\left(A \right)}_{\rho, \lambda}\left(p\right)=\bar{u}_{\rho}\left(
b_{A}\right)u_{\lambda}\left(p\right)=
 -\lambda \tilde{W}_{m_p}\left(-\lambda\;\rho\;\mathrm{p}\right)
 {D}^{*\;1/2}_{A \rho/2,-\lambda/2}\left(\varphi _p\;,\theta_p\;,-\varphi
_p\right) \; \label{mbstp26x}
\end{eqnarray}
with
\begin{equation}\label{wfunction}
\tilde{W}_{m_p}\left(\pm
\mathrm{p}\right)=\sqrt{\omega_{m_p}\left(\mathrm{p}\right)\pm \mathrm{p}}\;.
\end{equation}

The $\Gamma$ function is calculated in terms of the 4-vectors of the isotropic
tetrad $b_{A}$ and $n_{\lambda} ~~ (A, \lambda = \pm 1)$ using the basic relations
\texttt {MBS} (see appendix  \ref{app1}):
\begin{eqnarray}
&& \Gamma^{C,\;A}_{\rho,\;\sigma }\left[\gamma^{\mu}\right]=2\; \delta_{\sigma,\;
-\rho} \left(\delta_{C,\;-A}\; {b}_{-A}^{\mu}+A\;\delta_{C,A} \; {n}_{-A \times
\rho }^{\mu}\right)\;. \label{mbstp40b}
\end{eqnarray}

Since the scalar products of the isotropic tetrad vectors satisfy the relations
\begin{equation}
\left(b_\rho b_{-\lambda }\right)=\delta _{\lambda ,\;\rho }/2~~,~\left(n_\lambda
n_{-\rho }\right)=\delta _{\lambda ,\;\rho }/2~, ~\left( b_\rho n_\lambda \right)
=0\;, \label{min4}
\end{equation}
using the expansion coefficients (\ref{mbstp26x}), we find the product of fermionic
currents in  Eq.(\ref{vector1}) in the form
\begin{eqnarray}
&& N_{\mathrm{k},\mathrm{k^{\prime}}} \bar{u}_{\lambda _{p_1}}\left(p_{1}\right)
\gamma^{\mu} u_{\lambda _{k_1}}\left(k_{1}\right) \bar{u}_{\lambda
_{p_2}}\left(p_{2}\right)\gamma_{\mu} u_{\lambda
_{k_2}}\left(k_{2}\right)=\nonumber \\
&&\hspace{-1.2cm}= 2\sum_{\sigma,\rho = -1}^1 \sqrt{\left(1-\sigma \lambda
_{k_1}\mathrm{\upsilon}_{k_1}\right)\left(1-\rho\lambda
_{k_2}\mathrm{\upsilon}_{k_2}\right)} \sqrt{\left(1-\sigma \lambda
_{p_1}\mathrm{\upsilon}_{p_1}\right)\left(1-\rho\lambda
_{p_2}\mathrm{\upsilon}_{p_2}\right)}\times
\nonumber \\
&& \times \bigg[ \delta _{\lambda _{k_{1}},\lambda _{k_{2}}}\rho\;\sigma
D_{-\lambda_{k_{1}}/2,\lambda _{p_{1}}/2}^{\ast\;1/2}\left(\varphi,\beta,-\varphi
\right) D_{\lambda_{k_{1}}/2,-\lambda
_{p_{2}}/2}^{\ast\;1/2}\left(\varphi,\beta,-\varphi \right)+\nonumber \\
&& + \delta _{\rho \lambda _{k_{1}},\sigma \lambda _{k_{2}}}
D_{\lambda_{k_{1}}/2,\lambda _{p_{1}}/2}^{\ast\;1/2}\left(\varphi,\beta,-\varphi
\right) D_{-\lambda_{k_{2}}/2,-\lambda
_{p_{2}}/2}^{\ast\;1/2}\left(\varphi,\beta,-\varphi \right) \bigg] \;.
\label{kgplxy}
\end{eqnarray}

The following notation is introduced in the equation (\ref{kgplxy}):
\begin{equation}
\label{dopf1} \mathrm{\upsilon}_{k_1}=
\frac{\mathrm{k}}{\omega_{m_1}\left(\mathrm{k}\right)}\;,
\;\;\mathrm{\upsilon}_{p_1}=
\frac{\mathrm{k}^{\prime}}{\omega_{m_1}\left(\mathrm{k}^{\prime}\right)}\;,\;\;
\mathrm{\upsilon}_{k_2}=
\frac{\mathrm{k}}{\omega_{m_2}\left(\mathrm{k}\right)}\;,\;\;
\mathrm{\upsilon}_{p_2}=
\frac{\mathrm{k}^{\prime}}{\omega_{m_2}\left(\mathrm{k}^{\prime}\right)}\;.
\end{equation}

Further, according to (\ref{hel2}), for the integration over angular variables the
spinor structures of the potential (\ ref {vector1}) multiplied by the function
$D_{\lambda,\lambda^{\prime}}^{\;J} \left(\varphi,\beta,-\varphi \right)$ with
$\lambda=\left(\lambda_{k_{1}}-\lambda_{k_{2}}\right)/2$  and
$\lambda^{\prime}=\left(\lambda_{p_{1}}-\lambda_{p_{2}}\right)/2$.

The representation of the spinor part in the form (\ref{kgplxy}) and the
Clebsch-Gordan expansion for $D$-matrices allow us to write the formula
(\ref{hel2}) into the integrand, as a linear combination of the Legendre
polynomials $\mathrm{P}_{\ell}\left(\cos \beta \right)$, and thus separate the
angular variables and the
$\mathrm{k}=\left|\mathbf{k}\right|,\mathrm{k}^{\prime}=\left|\mathbf{k}^{\prime}\right|$.
This construction greatly simplifies the next stage of integration. To shorten the
notation of the calculated structures, we introduce auxiliary functions
\begin{eqnarray}
&&G_{\lambda _{k_{1}},\lambda _{k_{2}};\lambda _{p_{1}},\lambda
_{p_{2}}}^{J,s_1,s_2}\left[ \Phi\left(x\right) \right]= \nonumber \\
&&\hspace{-1.2cm}= \sum_{s=\left\vert
s_1-s_2\right\vert}^{s_1+s_2}\sum_{\ell=\left\vert
J-s\right\vert}^{J+s}\frac{\left( 2\ell+1\right) }{\left( 2J+1\right) }\; {
\mathbf{C} \left\{ \smallmatrix
\hspace{-2pt} s_1 &\hspace{-2pt} s_2 &\hspace{-2pt} s \\
\hspace{-2pt} \lambda_{k_1}/2,&\hspace{-2pt} -\lambda_{k_2}/2,&\hspace{-2pt}
\lambda
\endsmallmatrix
\right\} } { \mathbf{C} \left\{ \smallmatrix
\hspace{-2pt} \ell &\hspace{-2pt} s &\hspace{-2pt} J \\
\hspace{-2pt} 0,&\hspace{-2pt} \lambda,&\hspace{-2pt} \lambda
\endsmallmatrix
\right\} } \times
\nonumber\\
&&\times { \mathbf{C} \left\{ \smallmatrix
\hspace{-2pt} s_1 &\hspace{-2pt} s_2 &\hspace{-2pt} s \\
\hspace{-2pt} \lambda_{p_1}/2,&\hspace{-2pt} -\lambda_{p_2}/2,&\hspace{-2pt}
\lambda^{\prime}
\endsmallmatrix
\right\} }{ \mathbf{C} \left\{ \smallmatrix
\hspace{-2pt} \ell &\hspace{-2pt} s &\hspace{-2pt} J \\
\hspace{-2pt} 0,&\hspace{-2pt} \lambda^{\prime},&\hspace{-2pt} \lambda^{\prime}
\endsmallmatrix
\right\} } \Phi_{\ell}\left(x\right)\; \label{gfunc}
\end{eqnarray}
and
\begin{equation*}
\label{wfunct}
W_{\lambda,\rho}\left(\mathrm{k}\right)=\sqrt{1+\lambda\;\mathrm{\upsilon}_{k_1}}
\sqrt{1+\rho\mathrm{\upsilon}_{k_2}}\;,\;\;
W_{\lambda,\rho}\left(\mathrm{k}^{\prime}\right)=
\sqrt{1+\lambda\;\mathrm{\upsilon}_{p_1}} \sqrt{1+\rho\mathrm{\upsilon}_{p_2}}\;.
\end{equation*}

Other spinor structures of the potential (\ref{vector1}) have a similar
(\ref{kgplxy}) structure. Let us present the final expressions for the spinor
structures multiplied by the function $D_{\lambda,\;\lambda^{\prime}}^{\;J}
\left(\varphi,\beta,-\varphi \right)$:
\begin{eqnarray}
&&\hspace{-15mm}D_{\lambda,\;\lambda^{\prime}}^{\;J} \left(\varphi,\beta,-\varphi
\right)\;N_{\mathrm{k},\mathrm{k^{\prime}}} \bar{u}_{\lambda
_{p_1}}\left(p_{1}\right) \gamma^{\mu} u_{\lambda _{k_1}}\left(k_{1}\right)
\bar{u}_{\lambda _{p_2}}\left(p_{2}\right)\gamma_{\mu} u_{\lambda
_{k_2}}\left(k_{2}\right)=\nonumber \\
&&\hspace{-1.2cm}= 2\sum_{\sigma,\rho = -1}^1 W_{-\sigma \lambda
_{k_1},\;-\rho\lambda _{k_2} } \left(\mathrm{k}\right)\; W_{-\sigma \lambda
_{p_1},-\rho\;\lambda _{p_2}}\left(\mathrm{k}^{\prime}\right)\times
\nonumber \\
&&\hspace{-17mm} \times \bigg[ \delta _{\lambda _{k_{1}},\lambda
_{k_{2}}}\rho\;\sigma G_{-\lambda _{k_{1}},\lambda _{k_{1}};\lambda
_{p_{1}},\lambda _{p_{2}}}^{J,1/2,1/2}\left[\mathrm{P}_{\ell}\left(x\right)
\right]+ \delta _{\rho \lambda _{k_{1}},\sigma \lambda _{k_{2}}} G_{\lambda
_{k_{1}},\lambda _{k_{2}};\lambda _{p_{1}},\lambda
_{p_{2}}}^{J,1/2,1/2}\left[\mathrm{P}_{\ell}\left(x\right)\right]\bigg]\;,
\label{kgp1x}
\end{eqnarray}
\begin{eqnarray}
&&\hspace{-1.4cm}D_{\lambda,\;\lambda^{\prime}}^{\;J} \left(\varphi,\beta,-\varphi
\right)\;N_{\mathrm{k},\mathrm{k^{\prime}}} \bar{u}_{\lambda
_{p_1}}\left(p_{1}\right) u_{\lambda _{k_1}}\left(k_{1}\right)\bar{u}_{\lambda
_{p_2}}\left(p_{2}\right)\left(\hat{p}_1+\hat{k}_1\right)  u_{\lambda
_{k_2}}\left(k_{2}\right)=\nonumber \\
&&\hspace{-1.2cm}= \sum_{\sigma,\rho = -1}^{1} W_{-\sigma \lambda
_{k_1},\;-\rho\lambda _{k_2} } \left(\mathrm{k}\right)\; W_{\sigma \lambda
_{p_1},-\rho\;\lambda _{p_2}}\left(\mathrm{k}^{\prime}\right) \times
\nonumber \\
&&\hspace{-1.4cm} \times \bigg[ \mathrm{k}^{\prime}\;\rho\;\lambda _{k_2}\left(3\;
G_{\lambda _{k_{1}},\lambda _{k_{2}};\lambda _{p_{1}},\lambda
_{p_{2}}}^{J,1/2,1/2}\left[x\;\mathrm{P}_{\ell}\left(x\right) \right] -2\;
G_{\lambda _{k_{1}},\lambda _{k_{2}};\lambda _{p_{1}},\lambda
_{p_{2}}}^{J,1/2,3/2}\left[\mathrm{P}_{\ell}\left(x\right) \right] \right)+
\nonumber \\
&& \hspace{-1.4cm}+ G_{\lambda _{k_{1}},\lambda _{k_{2}};\lambda _{p_{1}},\lambda
_{p_{2}}}^{J,1/2,1/2}\left[\mathrm{P}_{\ell}\left(x\right)
\right]\left\{\rho\left(3\lambda _{k_2}\mathrm{k}-2\lambda
_{p_2}\mathrm{k}^{\prime}\right)-3\left(\omega _{m_1}\left(\mathrm{k}\right)+
\omega_{m_1}\left(\mathrm{k}^{\prime}\right)\right) \right\}\bigg], \label{kgp2x}
\end{eqnarray}
\begin{eqnarray}
&&\hspace{-1.2cm}D_{\lambda,\;\lambda^{\prime}}^{\;J} \left(\varphi,\beta,-\varphi
\right)\;N_{\mathrm{k},\mathrm{k^{\prime}}} \bar{u}_{\lambda
_{p_1}}\left(p_{1}\right)\left(\hat{p}_2+\hat{k}_2\right)  u_{\lambda
_{k_1}}\left(k_{1}\right)\bar{u}_{\lambda _{p_2}}\left(p_{2}\right) u_{\lambda
_{k_2}}\left(k_{2}\right)=\nonumber \\
&&\hspace{-1.2cm}= \sum_{\sigma,\rho = -1}^1 W_{-\sigma \lambda
_{k_1},\;-\rho\lambda _{k_2} } \left(\mathrm{k}\right)\; W_{-\sigma \lambda
_{p_1},\rho\;\lambda _{p_2}}\left(\mathrm{k}^{\prime}\right)\times
\nonumber \\
&&\hspace{-1.4cm} \times \bigg[ \mathrm{k}^{\prime}\sigma\lambda _{k_1}\left(3\;
G_{\lambda _{k_{1}},\lambda _{k_{2}};\lambda _{p_{1}},\lambda
_{p_{2}}}^{J,1/2,1/2}\left[x\;\mathrm{P}_{\ell}\left(x\right) \right] -2\;
G_{\lambda _{k_{1}},\lambda _{k_{2}};\lambda _{p_{1}},\lambda
_{p_{2}}}^{J,3/2,1/2}\left[\mathrm{P}_{\ell}\left(x\right) \right]\right)+
\nonumber \\
&& \hspace{-1.4cm}+ G_{\lambda _{k_{1}},\lambda _{k_{2}};\lambda _{p_{1}},\lambda
_{p_{2}}}^{J,1/2,1/2}\left[\mathrm{P}_{\ell}\left(x\right) \right]\left\{\sigma
\left(3\lambda _{k_1}\mathrm{k}-2\lambda
_{p_1}\mathrm{k}^{\prime}\right)-3\left(\omega _{m_2}\left(\mathrm{k}\right)+
\omega_{m_2}\left(\mathrm{k}^{\prime}\right)\right)\right\}\bigg], \label{kgp3x}
\end{eqnarray}
\begin{eqnarray}
&&\hspace{-1.4cm}D_{\lambda,\;\lambda^{\prime}}^{\;J} \left(\varphi,\beta,-\varphi
\right)\;N_{\mathrm{k},\mathrm{k^{\prime}}}
\left(p_{1}+k_{1}\right)^{\nu}\left(p_{2}+k_{2}\right)_{\nu}\times\nonumber \\
&&\hspace{-1.2cm}\times\bar{u}_{\lambda _{p_1}}\left(p_{1}\right) u_{\lambda
_{k_1}}\left(k_{1}\right) \bar{u}_{\lambda
_{p_2}}\left(p_{2}\right) u_{\lambda _{k_2}}\left(k_{2}\right)=\nonumber \\
&&\hspace{-1.4cm}= \sum_{\sigma,\rho = -1}^1 W_{-\sigma \lambda
_{k_1},\;-\rho\lambda _{k_2} } \left(\mathrm{k}\right)\; W_{\sigma \lambda
_{p_1},\rho\;\lambda _{p_2}}\left(\mathrm{k}^{\prime}\right)\times
\nonumber \\
&&\hspace{-1.4cm} \times \bigg[
\left(\mathrm{k}^{\prime\;2}+\mathrm{k}^{2}+\left(\omega
_{m_1}\left(\mathrm{k}\right)+
\omega_{m_1}\left(\mathrm{k}^{\prime}\right)\right)\left(\omega
_{m_2}\left(\mathrm{k}\right)+ \omega_{m_2}\left(\mathrm{k}^{\prime}\right)\right)
\right) \times\nonumber \\
&& \hspace{-1.4cm}\times G_{\lambda _{k_{1}},\lambda _{k_{2}};\lambda
_{p_{1}},\lambda _{p_{2}}}^{J,1/2,1/2}\left[\mathrm{P}_{\ell}\left(x\right)
\right]+ 2 \mathrm{k}\; \mathrm{k}^{\prime} G_{\lambda _{k_{1}},\lambda
_{k_{2}};\lambda _{p_{1}},\lambda
_{p_{2}}}^{J,1/2,1/2}\left[x\;\mathrm{P}_{\ell}\left(x\right)\right]\bigg]\;,
\label{kgp4x}
\end{eqnarray}
where $x=\cos \beta$.

The second part of the potential, using the Eqs. (\ref{qfot}), (\ref{cal2}) and
(\ref{q1q2}), is converted into a product that contains zero current components:
\begin{eqnarray}
&& V_{\lambda_{p_1},\lambda_{p_2}\;
\lambda_{k_1},\lambda_{k_2}}^{\left(\mathrm{B}\right)}\left({\bf k}^{\prime}, {\bf
k} \right)= \frac{Z \alpha \hskip 1pt \Pi\left( \alpha,q^2\right)}{8 \pi^{2}
q^4}\left(\omega _{m_1}\left( \mathrm{k}^{\prime} \right)-\omega
_{m_1}\left(\mathrm{k}\right)
\right) \times\nonumber \\
&&\times \left(\omega _{m_{2}}\left(\mathrm{k}^{\prime}\right)-\omega
_{m_{2}}\left(\mathrm{k}\right)\right)\;N_{\mathrm{k},\mathrm{k^{\prime}}}\;J_{\lambda_{p_1},\lambda_{k_1}}^{\left(0\right)
}\left(p_1,k_1\right)\;J_{\lambda
_{p_2},\lambda_{k_2}}^{\left(0\right)}\left(p_2,k_2\right) \;.\label{potb1}
\end{eqnarray}

Applying the method of basis spinors for calculating the spinor part (\ref{potb1})
and the method described above, we find that
\begin{eqnarray}
&&  N_{\mathrm{k},\mathrm{k^{\prime}}} D_{\lambda,\;\lambda^{\prime}}^{\;J}
\left(\varphi,\beta,-\varphi
\right)\;N_{\mathrm{k},\mathrm{k^{\prime}}}\;J_{\lambda_{p_1},\lambda_{k_1}}^{\left(0\right)
}\left(p_1,k_1\right)\;J_{\lambda
_{p_2},\lambda_{k_2}}^{\left(0\right)}\left(p_2,k_2\right)=\nonumber \\
&&=\frac{1}{4 m_1\; m_2} \sum_{\sigma,\rho = -1}^1 W_{-\sigma \lambda
_{k_1},\;-\rho\lambda _{k_2} } \left(\mathrm{k}\right)G_{\lambda _{k_{1}},\lambda
_{k_{2}};\lambda _{p_{1}},\lambda
_{p_{2}}}^{J,1/2,1/2}\left[\mathrm{P}_{\ell}\left(x\right) \right]\times\nonumber \\
&&\hspace{-1.2cm}\times\bigg[4 m_1\;m_2\;K^{\mathrm{\left(I\right)
}}\left(\tilde{q}^2\right) W_{-\sigma \lambda _{p_1},-\rho\;\lambda
_{p_2}}\left(\mathrm{k}^{\prime}\right) -2 m_2\; K^{\mathrm{\left(II\right)
}}\left(\tilde{q}^2\right) W_{\sigma \lambda
_{p_1},-\rho\;\lambda _{p_2}}\left(\mathrm{k}^{\prime}\right)\times\nonumber \\
&&\times \left(\omega _{m_1}\left( \mathrm{k} \right)+\omega
_{m_1}\left(\mathrm{k}^{\prime}\right)\right)-2 m_1\; K^{\mathrm{\left(III\right)
}}\left(\tilde{q}^2\right) W_{-\sigma \lambda _{p_1},\rho\;\lambda
_{p_2}}\left(\mathrm{k}^{\prime}\right)\times\nonumber \\
&&\times \left(\omega _{m_{2}}\left( \mathrm{k}^{\prime} \right)+\omega
_{m_{2}}\left(\mathrm{k}\right)\right)+K^{\mathrm{\left(IV\right)
}}\left(\tilde{q}^2\right) W_{\sigma \lambda _{p_1},\rho\;\lambda
_{p_2}}\left(\mathrm{k}^{\prime}\right)\times\nonumber \\
&&\hspace{-1.2cm}\times \left(\omega _{m_{2}}\left( \mathrm{k}^{\prime}
\right)+\omega _{m_{2}}\left(\mathrm{k}\right)\right) \left(\omega _{m_1}\left(
\mathrm{k} \right)+\omega _{m_1}\left(\mathrm{k}^{\prime}\right)\right)\bigg] \;
\;,\label{kgp5x}
\end{eqnarray}
where the functions $K\left(\tilde{q}^2\right)$  are defined by equations
(\ref{funk1})-(\ref{funk4}).

\section{Radial equation kernel structure}

The main characteristic of our calculation method is to use the momentum space and
accurate relativistic evaluation of radial kernel $V^{J}_{\ell,S\; ;\ell^{\prime},
S^{\prime}}\left(\mathrm{k},\mathrm{k^{\prime}}\right)$. After accurate analytic
calculation of potential spinor part (\ref{va13a}) with the help of the method of basis
spinors radial kernel of the relativistic fermion-fermion system
$V^{J}_{\ell^{\prime}, S^{\prime},\ell,
S}\left(\mathrm{k^{\prime}},\mathrm{k}\right)$ with arbitrary angular momentum $J$
and spin total momentum $S=0,1$ is obtained using the equation (\ref{hel2}).

\begin{eqnarray}
&& V^{J}_{\ell^{\prime},S^{\prime}\;
;\ell,S}\left(\mathrm{k^{\prime}},\mathrm{{k}}\right)=
\frac{\sqrt{\left(2\ell+1\right) \left(2\ell^{\prime}+1\right)
}}{2J+1}\left(-1\right) \frac{Z\;\alpha}{4 \pi} \sum_{\lambda_{k_{1,2}},
\lambda_{p_{1,2}}=-1}^{1} {\mathbf{C} \left\{ \smallmatrix
\hspace{-2pt} 1/2 &\hspace{-2pt} 1/2 &\hspace{-2pt} S \\
\hspace{-2pt} \lambda_{k_1}/2,&\hspace{-2pt} -\lambda_{k_2}/2,&\hspace{-2pt}
\lambda
\endsmallmatrix
\right\}}\times\nonumber\\
&& \times  {\mathbf{C} \left\{ \smallmatrix
\hspace{-2pt}{\ell} &\hspace{-2pt} S &\hspace{-2pt} J \\
\hspace{-2pt} 0 ,&\hspace{-2pt}\lambda,&\hspace{-2pt} \lambda
\endsmallmatrix
\right\}} {\mathbf{C} \left\{ \smallmatrix
\hspace{-2pt}{\ell^{\prime}} &\hspace{-2pt} S^{\prime} &\hspace{-2pt} J^{\prime} \\
\hspace{-2pt} 0 ,&\hspace{-2pt}\lambda^{\prime},&\hspace{-2pt} \lambda^{\prime}
\endsmallmatrix
\right\}} {\mathbf{C} \left\{ \smallmatrix
\hspace{-2pt} 1/2 &\hspace{-2pt} 1/2 &\hspace{-2pt} S \\
\hspace{-2pt} \lambda_{p_1}/2,&\hspace{-2pt} -\lambda_{p_2}/2,&\hspace{-2pt}
\lambda
\endsmallmatrix
\right\}} \left. \Big(\right. V_{\lambda_{k_{1}},\lambda_{k_{2}}, \lambda_{p_{1}},
\lambda_{p_{2}}}^{\mathrm{I}}+V_{\lambda_{k_{1}},\lambda_{k_{2}}, \lambda_{p_{1}},
\lambda_{p_{2}}}^{\mathrm{II}}+\nonumber\\
&&\hspace{-1.2cm}+V_{\lambda_{k_{1}},\lambda_{k_{2}}, \lambda_{p_{1}},
\lambda_{p_{2}}}^{\mathrm{III}}+V_{\lambda_{k_{1}},\lambda_{k_{2}},
\lambda_{p_{1}}, \lambda_{p_{2}}}^{\mathrm{IV}}+ V_{\lambda_{p_1},\lambda_{p_2}\;
\lambda_{k_1},\lambda_{k_2}}^{\left(\mathrm{B}\right)}\left.\right.\Big)\;,
\label{redmat1x}
\end{eqnarray}
where
\begin{eqnarray}
&&V_{\lambda_{k_{1}},\lambda_{k_{2}}, \lambda_{p_{1}},
\lambda_{p_{2}}}^{\mathrm{I}}= 2\sum_{\sigma,\rho = -1}^1 W_{-\sigma \lambda
_{k_1},\;-\rho\lambda _{k_2} } \left(\mathrm{k}\right)\; W_{-\sigma \lambda
_{p_1},-\rho\;\lambda _{p_2}}\left(\mathrm{k}^{\prime}\right) \bigg[ \delta
_{\lambda _{k_{1}},\lambda _{k_{2}}}\rho\;\sigma \times
\nonumber \\
&& \times G_{-\lambda _{k_{1}},\lambda _{k_{1}};\lambda _{p_{1}},\lambda
_{p_{2}}}^{J,1/2,1/2}\left[\mathrm{\tilde{R}}_{\ell}^{\left(I\right)
}\left(\mathrm{k},\mathrm{k}^{\prime}\right) \right]+ \delta _{\rho \lambda
_{k_{1}},\sigma \lambda _{k_{2}}} G_{\lambda _{k_{1}},\lambda _{k_{2}};\lambda
_{p_{1}},\lambda
_{p_{2}}}^{J,1/2,1/2}\left[\mathrm{\tilde{R}}_{\ell}^{\left(I\right)
}\left(\mathrm{k},\mathrm{k}^{\prime}\right) \right] \bigg]\;, \label{v1x}
\end{eqnarray}
\begin{eqnarray}
&&V_{\lambda_{k_{1}},\lambda_{k_{2}}, \lambda_{p_{1}},
\lambda_{p_{2}}}^{\mathrm{II}}= -\;\frac{1}{2\;m_{1}}\sum_{\sigma,\rho = -1}^{1}
W_{-\sigma \lambda _{k_1},\;-\rho\lambda _{k_2} } \left(\mathrm{k}\right)\;
W_{\sigma \lambda _{p_1},-\rho\;\lambda _{p_2}}\left(\mathrm{k}^{\prime}\right)
\bigg[ \mathrm{k}^{\prime}\;\rho\;\lambda _{k_2}\times
\nonumber \\
&& \times\left(3\; G_{\lambda _{k_{1}},\lambda _{k_{2}};\lambda _{p_{1}},\lambda
_{p_{2}}}^{J,1/2,1/2}\left[\tilde{Z}^{\left(\mathrm{II}\right)
}\left({\mathrm{k}}^{\prime},\mathrm{k}\right) \right] -2\; G_{\lambda
_{k_{1}},\lambda _{k_{2}};\lambda _{p_{1}},\lambda
_{p_{2}}}^{J,1/2,3/2}\left[\tilde{R}^{\left(\mathrm{II}\right)
}\left({\mathrm{k}}^{\prime},\mathrm{k}\right) \right] \right)+
\nonumber \\
&& + G_{\lambda _{k_{1}},\lambda _{k_{2}};\lambda _{p_{1}},\lambda
_{p_{2}}}^{J,1/2,1/2}\left[\tilde{R}^{\left(\mathrm{II}\right)
}\left({\mathrm{k}}^{\prime},\mathrm{k}\right) \right]\left\{\rho\left(3\lambda
_{k_2}\mathrm{k}-2\lambda _{p_2}\mathrm{k}^{\prime}\right)-3\left(\omega
_{m_1}\left(\mathrm{k}\right)+ \omega_{m_1}\left(\mathrm{k}^{\prime}\right)\right)
\right\}\bigg] \;,\nonumber \\ \label{v3x}
\end{eqnarray}
\begin{eqnarray}
&&V_{\lambda_{k_{1}},\lambda_{k_{2}}, \lambda_{p_{1}},
\lambda_{p_{2}}}^{\mathrm{III}}=-\;\frac{1}{2\;m_{2}}\sum_{\sigma,\rho = -1}^1
W_{-\sigma \lambda _{k_1},\;-\rho\lambda _{k_2} } \left(\mathrm{k}\right)\;
W_{-\sigma \lambda _{p_1},\rho\;\lambda
_{p_2}}\left(\mathrm{k}^{\prime}\right)\times
\nonumber \\
&& \times \bigg[ \mathrm{k}^{\prime}\sigma\lambda _{k_1}\left(3\; G_{\lambda
_{k_{1}},\lambda _{k_{2}};\lambda _{p_{1}},\lambda
_{p_{2}}}^{J,1/2,1/2}\left[\tilde{Z}^{\left(\mathrm{III}\right)
}\left({\mathrm{k}}^{\prime},\mathrm{k}\right) \right] -2\; G_{\lambda
_{k_{1}},\lambda _{k_{2}};\lambda _{p_{1}},\lambda
_{p_{2}}}^{J,3/2,1/2}\left[\tilde{R}^{\left(\mathrm{III}\right)
}\left({\mathrm{k}}^{\prime},\mathrm{k}\right) \right]\right)+
\nonumber \\
&& + G_{\lambda _{k_{1}},\lambda _{k_{2}};\lambda _{p_{1}},\lambda
_{p_{2}}}^{J,1/2,1/2}\left[\tilde{R}^{\left(\mathrm{III}\right)
}\left({\mathrm{k}}^{\prime},\mathrm{k}\right) \right]\left\{\sigma \left(3\lambda
_{k_1}\mathrm{k}-2\lambda _{p_1}\mathrm{k}^{\prime}\right)-3\left(\omega
_{m_2}\left(\mathrm{k}\right)+ \omega_{m_2}\left(\mathrm{k}^{\prime}\right)\right)
\right\}  \bigg]\;,\nonumber \\ \label{v4x}
\end{eqnarray}
\begin{eqnarray}
&&V_{\lambda_{k_{1}},\lambda_{k_{2}}, \lambda_{p_{1}},
\lambda_{p_{2}}}^{\mathrm{IV}}= \frac{1}{4\;m_{1}\;m_{2}}\sum_{\sigma,\rho = -1}^1
W_{-\sigma \lambda _{k_1},\;-\rho\lambda _{k_2} } \left(\mathrm{k}\right)\;
W_{\sigma \lambda _{p_1},\rho\;\lambda
_{p_2}}\left(\mathrm{k}^{\prime}\right)\times
\nonumber \\
&& \times \bigg[ \left(\mathrm{k}^{\prime\;2}+\mathrm{k}^{2}+\left(\omega
_{m_1}\left(\mathrm{k}\right)+
\omega_{m_1}\left(\mathrm{k}^{\prime}\right)\right)\left(\omega
_{m_2}\left(\mathrm{k}\right)+ \omega_{m_2}\left(\mathrm{k}^{\prime}\right)\right)
\right) \times\nonumber \\
&& \times G_{\lambda _{k_{1}},\lambda _{k_{2}};\lambda _{p_{1}},\lambda
_{p_{2}}}^{J,1/2,1/2}\left[\tilde{R}^{\left(\mathrm{IV}\right)
}\left({\mathrm{k}}^{\prime},\mathrm{k}\right)  \right]+ 2\; \mathrm{k}\;
\mathrm{k}^{\prime} G_{\lambda _{k_{1}},\lambda _{k_{2}};\lambda _{p_{1}},\lambda
_{p_{2}}}^{J,1/2,1/2}\left[\tilde{Z}^{\left(\mathrm{IV}\right)
}\left({\mathrm{k}}^{\prime},\mathrm{k}\right)  \right] \bigg] \;. \label{v4xz}
\end{eqnarray}
and
\begin{equation*}
\label{pertb2bx} \tilde{Z}_{\ell}\left({\mathrm{k}}^{\prime},\mathrm{k}\right)=
\frac{1}{2\ell+1}\;\left[\left(\ell+1\right)
\tilde{R}_{\ell+1}\left({\mathrm{k}}^{\prime},\mathrm{k}\right)+
\ell\;\tilde{R}_{\ell-1}\left({\mathrm{k}}^{\prime},\mathrm{k}\right)\right]\;,
\end{equation*}
\begin{eqnarray}
&&
W_{\lambda,\rho}\left(\mathrm{k}\right)=\sqrt{1+\lambda\;\mathrm{\upsilon}_{k_1}}
\sqrt{1+\rho\mathrm{\upsilon}_{k_2}}\;,
W_{\lambda,\rho}\left(\mathrm{k}^{\prime}\right)=
\sqrt{1+\lambda\;\mathrm{\upsilon}_{p_1}} \sqrt{1+\rho\mathrm{\upsilon}_{p_2}}
\label{wfunct1}
\end{eqnarray}
with
\begin{equation}
\label{dopf11} \mathrm{\upsilon}_{k_1}=
\frac{\mathrm{k}}{\omega_{m_1}\left(\mathrm{k}\right)}\;,
\;\;\mathrm{\upsilon}_{p_1}=
\frac{\mathrm{k}^{\prime}}{\omega_{m_1}\left(\mathrm{k}^{\prime}\right)}\;,\;\;
\mathrm{\upsilon}_{k_2}=
\frac{\mathrm{k}}{\omega_{m_2}\left(\mathrm{k}\right)}\;,\;\;
\mathrm{\upsilon}_{p_2}=
\frac{\mathrm{k}^{\prime}}{\omega_{m_2}\left(\mathrm{k}^{\prime}\right)}\;.
\end{equation}

The analytic expression of the last potential part is determined by
\begin{eqnarray}
&& \hspace{-1.0cm}V_{\lambda_{k_{1}},\lambda_{k_{2}}, \lambda_{p_{1}},
\lambda_{p_{2}}}^{\mathrm{B}}=  \sum_{\sigma,\rho = -1}^1 W_{-\sigma \lambda
_{k_1},\;-\rho\lambda _{k_2}}\left(\mathrm{k}\right)\bigg[G_{\lambda _{k_{1}},\lambda
_{k_{2}};\lambda _{p_{1}},\lambda
_{p_{2}}}^{J,1/2,1/2}\left[\tilde{U}^{\left(\mathrm{I}\right)
}\left({\mathrm{k}}^{\prime},\mathrm{k}\right) \right]\times\nonumber \\
&&\hspace{-1.0cm} \times W_{-\sigma \lambda
_{p_1},-\rho\;\lambda_{p_2}}\left(\mathrm{k}^{\prime}\right)-\frac{1}{2 m_2}\;
G_{\lambda _{k_{1}},\lambda _{k_{2}};\lambda _{p_{1}},\lambda
_{p_{2}}}^{J,1/2,1/2}\left[\tilde{U}^{\left(\mathrm{II}\right)
}\left({\mathrm{k}}^{\prime},\mathrm{k}\right) \right] \times\nonumber \\
&&\hspace{-1.0cm}\times W_{-\sigma \lambda _{p_1},\rho\;\lambda
_{p_2}}\left(\mathrm{k}^{\prime}\right)\left(\omega _{m_{2}}\left( \mathrm{k}^{\prime}
\right)+\omega _{m_{2}}\left(\mathrm{k}\right)\right)\bigg] \;.\label{v5x}
\end{eqnarray}

The functions $\tilde{R}_{\ell}\left({\mathrm{k}}^{\prime},\mathrm{k}\right)$ and
$\tilde{U}_{\ell}\left({\mathrm{k}}^{\prime},\mathrm{k}\right)$ in equations
(\ref{v1x})-(\ref{v5x}) are represented as the integrals
\begin{eqnarray}
&& \tilde{R}_{\ell}\left({\mathrm{k}}^{\prime},\mathrm{k}\right)=
\int\limits_{-1}^{1}\frac{K\left(\tilde{q}^2\right) \mathrm{P}_{\ell}\left(x \right)
}{q^{2}}\;\mathrm{d}x\;,\label{pertb3a}
\end{eqnarray}
\begin{eqnarray}
&& \tilde{U}_{\ell}\left({\mathrm{k}}^{\prime},\mathrm{k}\right)=
\varrho_{12}\left({\mathrm{k}}^{\prime},\mathrm{k}\right)\;
\int\limits_{-1}^{1}\frac{K\left(\tilde{q}^2\right)
\mathrm{P}_{\ell}\left(x \right)
}{q^{4}}\;\mathrm{d}x\;,\label{pertb3abc}
\end{eqnarray}
where the dimension factor $\varrho_{12}\left({\mathrm{k}}^{\prime},\mathrm{k}\right)$ is
\begin{equation}\label{varrho}
\varrho_{12}\left({\mathrm{k}}^{\prime},\mathrm{k}\right)=\left(\omega _{m_1}\left(
\mathrm{k}^{\prime} \right)-\omega _{m_1}\left(\mathrm{k}\right)
\right)\left(\omega _{m_{2}}\left( \mathrm{k} \right)-\omega
_{m_{2}}\left(\mathrm{k}^{\prime}\right)\right)\;
\end{equation}
and
\begin{equation}
\label{q2x} q^2=-2\mathrm{k}\;\mathrm{k}^{\prime}\left(y-x\right)\;,
\end{equation}
\begin{equation}\label{yf}
y=\frac{\mathrm{k}^2+{\mathrm{k}^{\prime}}^2}{2 \mathrm{k}\;\mathrm{k}^{\prime}}\;.
\end{equation}

In the case when $m_1= m_2$ the Lorentz structures $\gamma^{\mu}
\otimes \gamma_{\mu}$ and $I\otimes I$ of the potential (\ref{redmat1x}) coincide
with similar structures obtained in \cite{Brown79engl}.

Expanding in fermion velocities, it can be shown that the potential
(\ref{redmat1x}) transforms into both the nonrelativistic Schr\"{o}dinger equation
and the Breit equation in the momentum representation.

The potential in the form (\ref{redmat1x}) with the terms (\ref{v1x})-(\ref{v5x})
allows one to estimate the contributions of both the proton structure and the higher
order electromagnetic corrections. To calculate this or that correction it is
enough to define the explicit form of the function $\tilde {R}_{\ell}$ and $ \tilde
{U}_ {\ell}$, while the rest of the structure remains unchanged.

\section{Summary}

In this model we have the gauge invariant effective potential and the exact calculation
of the relativistic kernel of the two-fermion equation with electromagnetic interaction
that was performed using the method of basis spinors.

The resulting kernel of the radial equation (\ref {maineq1}) for an arbitrary
total angular momentum $J$ (total spin momentum $S =0,1 $) automatically takes into
account recoil effects and allows one to take into account higher-order
relativistic effects caused by the motion of fermions when calculating the energy
contributions.

The proposed technique can also be applied to build the potential of one-gluon
exchange without significant additional calculations.

\newpage
\appendix

\section{The Method of Basic Spinors \label{app1}}

When evaluating a Feynman amplitude involving fermions, the amplitude is expressed
as sum of terms which have the form
\begin{eqnarray}
&&\hspace{-3mm} \mathcal{M}_{\lambda _p,\lambda _k}\left(p,s_p,\; k,s_k\;
;Q\right)= \mathcal{M}_{\lambda _p,\lambda _k}\left(\left[p\right], \left[k\right]
;Q\right)=\nonumber\\
&&=\bar{w}^{A}_{\lambda _p}\left( p,s_p\right) Q ~w^{B}_{\lambda
_k}\left(k,s_k\right)\;, \label{anpic1}
\end{eqnarray}
where $\lambda_{p}$ and $\lambda_{k}$ are spin indices of the external fermions
with four-momenta $p,k$ and arbitrary polarization vectors $s_p,s_k$. The operator
$Q$ is a sum of products of Dirac $\gamma$-matrices. The notation $w^{A}_{\lambda
_p}\left(p,s_p\right)$ stands for either $u_{\lambda _p}\left( p,s_p\right)$
(bispinor of fermion; $A=+1$) or $\upsilon_{\lambda _p}\left( p,s_p\right)$
(bispinor of antifermion; $A=-1$).

The main aim of the calculation is to transform (\ref{anpic1}) into and explicitly scalar
form (scalar products of four-vectors, Lorentz tensors, and so on). The main
approach which has gained popularity in the past decades is to calculate Feynman
amplitudes directly. Many different methods of calculating reaction amplitudes with
fermions have been developed
\cite{Bellomo61,Bogush1962a,Gastmans:1990xh,Dittmaier:1998nn} et.al. In the paper
we describe an approach to Feynman diagrams which is based on the utilization of an
isotropic tetrad in the Minkowski space and massless basis spinors connected with it
and which we call the Method of Basis Spinors (\texttt{MBS})
\cite{Andreev2003ce,Andreev2009z}) Let us briefly describe the main
relationships of the $\mathbf{MBS}$.

\subsection{The isotropic tetrad and massless basis spinors}

Let us introduce the orthonormal four-vector basis in the Minkowski space which
satisfies the relations:
\begin{equation}
\label{pic1} l_0^{\mu}  l_0^{\nu} - \sum_{j=1}^{3}l_j^{\mu} l_j^{\nu} = g^{\mu
\nu}, ~~\left(l_{A} \cdot l_{B}\right)=g_{A B}\;,
\end{equation}
where $g$ is the Lorentz metric tensor.

With the help of vectors $l_{A}$ we can define lightlike vectors which form the
isotropic tetrad in the Minkowski space
\begin{equation}
b_\rho =(l_0+\rho l_3)/2 , \; n_\lambda =(\lambda l_1+\mathrm{i} l_2)/2\;,
~~~\left(\lambda,\rho= \pm 1\right). \label{pic2}
\end{equation}
From Eqs. (\ref{pic1}), (\ref{pic2}) it follows that
\begin{eqnarray}
&& \hspace{-7mm}(b_\rho \cdot b_{-\lambda })=(n_\rho \cdot n_{-\lambda
})=\frac{\delta _{\lambda ,\rho }}{2}\;,~\left(b_\rho \cdot n_\lambda \right) =0\;,
\label{pic3}
\end{eqnarray}
\begin{eqnarray}
&&g^{\mu \nu}= \sum_{\lambda =-1}^1\left[ \tilde{b}_\lambda ^\mu \cdot
b_{-\lambda}^\nu
+ \tilde{n}_\lambda ^\mu \cdot n_{-\lambda }^\nu \right] \;,\label{pic4} \\
&& \tilde{b}_\lambda ^\mu= 2 \;{b}_\lambda ^\mu\;, \;\tilde{n}_\lambda ^\mu = 2
\;{n}_\lambda ^\mu \;.
\end{eqnarray}

It is always possible to construct the basis of the isotropic tetrad (\ref{pic2}) as
numerical four-vectors
\begin{equation}\label{pic4a}
\left(b_{\pm 1}\right)_{\mu}=\left(1/2\right)\left\{1, 0, 0, \pm 1\right\}\;,\;
\left(n_{\pm 1}\right)_{\mu}=\left(1/2\right)\left\{0, \pm 1, \mathrm{i}, 0\right\}
\end{equation}
or by means of physical vectors for reaction.

By means of the isotropic tetrad (\ref{pic2}) we define {\it basis spinors}
$u_{\lambda}\left(b_{-1}\right)$ and $u_{\lambda}\left(b_{1}\right)$\;\;:
\begin{equation}
\slash{b}_{-1} u_\lambda \left( b_{-1}\right) =0\;,\;\;\; ~~u_\lambda
\left(b_{1}\right) \equiv
\slash{b}_{1}u_{-\lambda}\left(b_{-1}\right)\;,\label{pic7}
\end{equation}
\begin{equation}
\omega _\lambda u_\lambda \left(b_{A}\right)= u_\lambda
\left(b_{A}\right)\;,\;\;(A=\pm 1) \label{pic8}
\end{equation}
with matrix $\omega _{\lambda} = 1/2 \hskip 1pt \left( 1+\lambda \gamma_5\right)$
and normalization condition
\begin{equation}\label{pic8a}
u_\lambda \left( b_{A}\right) \bar{u}_\lambda \left(b_{A}\right) =\omega _\lambda
\slash{b}_{A}\;.
\end{equation}
The relative phase between basis spinors with different helicity is given by
\begin{equation}
\slash{n}_\lambda u_{-\rho }\left( b_{-1}\right) =\delta_{\lambda, \rho} u_\lambda
\left(b_{-1}\right)\;. \label{pic9}
\end{equation}

The important property of basis spinors (\ref{pic7}) is the \textbf{completeness
relation}:
\begin{equation}
\sum_{\lambda,A=-1}^{1} u_\lambda \left( b_A\right) \bar{u}_{-\lambda}
\left(b_{-A}\right)= \mathrm{I}\;, \label{pic10}
\end{equation}
which follows from Eqs.(\ref{pic7})--(\ref{pic9}). Thus, an arbitrary bispinor can
be decomposed in terms of basis spinors $u_{\lambda}\left(b_{A} \right)$.

\subsection{Main equations of the \texttt{MBS} and Dirac spinors}

An arbitrary Dirac spinor can be determined through the basis spinor (\ref{pic7})
with the help of projection operators
$\tau_{\lambda}\left(p,s_p\right)=u_{\lambda_{p}}
\left(p,s_p\right)\bar{u}_{\lambda_{p}} \left(p,s_p\right)$. Dirac spinors
$w^A_\lambda \left(p,s_p\right)$ for massive fermion and antifermion with
four-momentum $p\;( p^2=m_p^2 )$ , arbitrary polarization vector $s_p$ and spin
number $\lambda= \pm 1$  can be obtained with the help of basis spinors by means of
equation:
\begin{eqnarray}
&&\hspace{-7mm} w^A_\lambda \left(p,s_p\right) =\left(A \lambda\right)
\frac{\left(\slash{p}+A m_p\right) \left(1+ \lambda
\gamma_{5}\slash{s}_{p}\right)}{2\sqrt{\left({b}_{1}\cdot \left( p+m_p s_p
\right)\right)}}\; u_{-A \times \lambda} \left({b}_{1} \right) \label{anpic8}
\end{eqnarray}

Spinor products of basis spinors are simple and similar to scalar products of
isotropic tetrad vectors
\begin{equation}
\bar{u}_\lambda \left( b_C\right) u_{\rho}\left( b_A\right) =
\delta_{\lambda,-\rho} \delta_{C,-A}\;. \label{pic12}
\end{equation}

With the help of Eq.(\ref{pic4}) the Dirac matrix $\gamma^\mu $ can be rewritten as
\begin{equation}
\gamma ^\mu =\sum_{\lambda =-1}^1\left[\slash{b}_{-\lambda} \tilde{b}_\lambda ^\mu
+\slash{n}_{-\lambda } \tilde{n}_\lambda ^\mu \right]\; \label{pic5}
\end{equation}
and using Eqs.(\ref{pic8}),(\ref{pic9}) and (\ref{pic5}) we can obtain that
\begin{equation}
\hspace{-3mm}\gamma ^\mu\;u_\lambda \left( b_A\right) =\tilde{b}_A^\mu u_{-\lambda
}\left( b_{-A}\right) -A\; \tilde{n}_{-A \times\lambda }^\mu u_{-\lambda }\left(
b_A\right)\;, \label{pic11}
\end{equation}
which allows to transform the Dirac matrix into some combination of isotropic tetrad
vectors in the basis spinor space and
\begin{equation}
\label{pic13a} \gamma_{5}\; u_\rho \left(b_A\right)= \rho \;u_\rho
\left(b_A\right)\;.
\end{equation}

Eqs. (\ref{pic12}), (\ref{pic11})  and (\ref{pic13a})  underly the method of
basis spinors (\texttt{MBS}).

\subsection{The \texttt{MBS} and the technique of ``building'' blocks}

The \textbf{basic idea of the Method of Basis Spinors} is to replace Dirac spinors in
Eq.(\ref{anpic1}) by massless basis spinors $u_{\lambda}\left(b_{\pm 1}\right)$
(Eq.(\ref{anpic8})) and to use only three Eqs. (\ref{pic12}), (\ref{pic11})  and
(\ref{pic13a}) to calculate the matrix element (\ref{anpic1}) in terms of scalar
functions.

Let us consider an important type of the matrix element (\ref{anpic1}) when $p=b_{-C}$
and $k=b_A$, i.e.
\begin{eqnarray}
&& \mathcal{M}_{\sigma,-\rho}\left(b_{C}\;,\;b_{-A}\; ; Q\right)\equiv
\Gamma^{C,\;A}_{\sigma ,\rho}\left[Q\right] = \bar{u}_{\sigma} \left( b_{C}\right)
Q \;u_{-\rho} \left( b_{-A}\right)\;. \label{pic13}
\end{eqnarray}
We call this type of matrix element as the \textbf{basic matrix element}. By means of
\textsf{MBS} relations (\ref{pic12}), (\ref{pic11})  and (\ref{pic13a}) it is easy
to calculate $\Gamma^{C,A}_{\sigma, \rho}$ in terms of isotropic tetrad
vectors.

With the help of the completeness relation (\ref{pic10}) the amplitude (\ref{anpic1})
can be expressed as combinations of the lower-order matrix elements (``building''
blocks)
\begin{eqnarray}
&&\mathcal{M}_{\lambda _p,\lambda _k}\left(p,s_p\; k,s_k;Q\right)= \sum_{A,\;C,
\sigma,\;\rho =-1}^1 \left\{\bar{w}^{D}_{\lambda
_p}\left(p,s_p\right)u_{-\sigma}\left(b_{-C}\right)\right\} \times \nonumber\\&&
\hspace{-5mm}\times \left\{\bar{u}_{\sigma}\left(b_{C}\right)Q u_{-\rho}\left(
b_{-A}\right)\right\} \left\{\bar{u}_{\rho}\left( b_{A}\right) w^{F}_{\lambda
_k}\left(k,s_k\right)\right\}=
\nonumber\\
&& =\sum\limits_{\sigma,\; \rho =-1}^1 \sum_{A,\;C=-1}^1
\bar{s}^{\left(C,\;D\right)}_{\sigma,\lambda_{p}}\left(p,s_p\right)
\Gamma^{C,A}_{\sigma ,\rho }\left[Q \right] s^{\left(A,\;F\right)}_{\rho, \lambda
_k}\left(k,s_k\right)\;. \label{anpic2aac}
\end{eqnarray}
Decomposition coefficients for the helicity states of fermions can be easily
calculated:
\begin{equation}\label{s17}
s_{\rho,\lambda}^{(A,D)}(p, s_{\mathrm{hel}}) = D \lambda\;  W_{m}(-\lambda \rho D
\mathrm{p}) f(\rho\lambda,D) D_{A\rho/2,-D\lambda/2}^{* 1/2} (\phi,\theta,-\phi)
\end{equation}
where
\begin{eqnarray}\label{s18}
&&  W_{m}(\pm \mathrm{p})=\sqrt{\omega_{m}(\mathrm{p})\pm \mathrm{p}}\; , \; \; \;
\omega_{m}(\mathrm{p})=\sqrt{\mathrm{p}^{2}+m^{2}}\; , \; \;
\mathrm{p}=\left|\mathbf{p}\right|,
\nonumber\\
&&f(A,D) = \delta_{A,-1}+D\delta_{A,1}
\end{eqnarray}
and $D_{\sigma_{1},\sigma_{2}}^{1/2}(\phi,\theta,-\varphi)=\exp\left(-i\phi
\right)d_{\sigma_{1},\sigma_{2}}^{1/2}(\theta) \exp\left(-i\varphi \right)$ is the
Wigner function.


\begin{thebibliography}{0}%
\makeatletter
\providecommand \@ifxundefined [1]{%
 \@ifx{#1\undefined}
}%
\providecommand \@ifnum [1]{%
 \ifnum #1\expandafter \@firstoftwo
 \else \expandafter \@secondoftwo
 \fi
}%
\providecommand \@ifx [1]{%
 \ifx #1\expandafter \@firstoftwo
 \else \expandafter \@secondoftwo
 \fi
}%
\providecommand \natexlab [1]{#1}%
\providecommand \enquote  [1]{``#1''}%
\providecommand \bibnamefont  [1]{#1}%
\providecommand \bibfnamefont [1]{#1}%
\providecommand \citenamefont [1]{#1}%
\providecommand \href@noop [0]{\@secondoftwo}%
\providecommand \href [0]{\begingroup \@sanitize@url \@href}%
\providecommand \@href[1]{\@@startlink{#1}\@@href}%
\providecommand \@@href[1]{\endgroup#1\@@endlink}%
\providecommand \@sanitize@url [0]{\catcode `\\12\catcode `\$12\catcode
  `\&12\catcode `\#12\catcode `\^12\catcode `\_12\catcode `\%12\relax}%
\providecommand \@@startlink[1]{}%
\providecommand \@@endlink[0]{}%
\providecommand \url  [0]{\begingroup\@sanitize@url \@url }%
\providecommand \@url [1]{\endgroup\@href {#1}{\urlprefix }}%
\providecommand \urlprefix  [0]{URL }%
\providecommand \Eprint [0]{\href }%
\providecommand \doibase [0]{http://dx.doi.org/}%
\providecommand \selectlanguage [0]{\@gobble}%
\providecommand \bibinfo  [0]{\@secondoftwo}%
\providecommand \bibfield  [0]{\@secondoftwo}%
\providecommand \translation [1]{[#1]}%
\providecommand \BibitemOpen [0]{}%
\providecommand \bibitemStop [0]{}%
\providecommand \bibitemNoStop [0]{.\EOS\space}%
\providecommand \EOS [0]{\spacefactor3000\relax}%
\providecommand \BibitemShut  [1]{\csname bibitem#1\endcsname}%
\let\auto@bib@innerbib\@empty
\end{thebibliography}%


\newpage
\begin{thebibliography}{33}
\expandafter\ifx\csname natexlab\endcsname\relax\def\natexlab#1{#1}\fi
\expandafter\ifx\csname bibnamefont\endcsname\relax
  \def\bibnamefont#1{#1}\fi
\expandafter\ifx\csname bibfnamefont\endcsname\relax
  \def\bibfnamefont#1{#1}\fi
\expandafter\ifx\csname citenamefont\endcsname\relax
  \def\citenamefont#1{#1}\fi
\expandafter\ifx\csname url\endcsname\relax
  \def\url#1{\texttt{#1}}\fi
\expandafter\ifx\csname urlprefix\endcsname\relax\def\urlprefix{URL }\fi
\providecommand{\bibinfo}[2]{#2} \providecommand{\eprint}[2][]{\url{#2}}

\bibitem[{\citenamefont{Eides et~al.}(2001)\citenamefont{Eides, Grotch, and
  Shelyuto}}]{Eides:2000xc}
\bibinfo{author}{\bibfnamefont{M.~I.} \bibnamefont{Eides}},
  \bibinfo{author}{\bibfnamefont{H.}~\bibnamefont{Grotch}}, \bibnamefont{and}
  \bibinfo{author}{\bibfnamefont{V.~A.} \bibnamefont{Shelyuto}},
  \bibinfo{journal}{Phys. Rept.} \textbf{\bibinfo{volume}{342}},
  \bibinfo{pages}{63} (\bibinfo{year}{2001}), \eprint{hep-ph/0002158}.

\bibitem[{\citenamefont{Karshenboim}(2004)}]{Karshenboim:2003vs}
\bibinfo{author}{\bibfnamefont{S.~G.} \bibnamefont{Karshenboim}},
  \bibinfo{journal}{Int. J. Mod. PHys.} \textbf{\bibinfo{volume}{A19}},
  \bibinfo{pages}{3879} (\bibinfo{year}{2004}), \eprint{hep-ph/0310099}.

\bibitem[{\citenamefont{Karshenboim et~al.}(2006)}]{Karshenboim:2006ht}
\bibinfo{author}{\bibfnamefont{S.~G.} \bibnamefont{Karshenboim}}
  \bibnamefont{et~al.}, \bibinfo{journal}{Nucl. Phys. Proc. Suppl.}
  \textbf{\bibinfo{volume}{162}}, \bibinfo{pages}{260} (\bibinfo{year}{2006}),
  \eprint{hep-ph/0608236}.

\bibitem[{\citenamefont{Pohl et~al.}(2010)}]{Pohl2010er}
\bibinfo{author}{\bibfnamefont{R.}~\bibnamefont{Pohl}} \bibnamefont{et~al.},
  \bibinfo{journal}{Nature} \textbf{\bibinfo{volume}{466}},
  \bibinfo{pages}{231} (\bibinfo{year}{2010}).

\bibitem[{\citenamefont{De~Rujula}(2010)}]{DeRujula:2010dp}
\bibinfo{author}{\bibfnamefont{A.}~\bibnamefont{De~Rujula}},
  \bibinfo{journal}{Phys.Lett.} \textbf{\bibinfo{volume}{B693}},
  \bibinfo{pages}{555} (\bibinfo{year}{2010}), \eprint{1008.3861}.

\bibitem[{\citenamefont{Jentschura}(2011{\natexlab{a}})}]{Jentschura2011500}
\bibinfo{author}{\bibfnamefont{U.}~\bibnamefont{Jentschura}},
  \bibinfo{journal}{Annals of Physics} \textbf{\bibinfo{volume}{326}},
  \bibinfo{pages}{500 } (\bibinfo{year}{2011}{\natexlab{a}}), ISSN
  \bibinfo{issn}{0003-4916}, \eprint{1011.5275},
  \urlprefix\url{http://www.sciencedirect.com/science/article/pii/S00034916100%
02010}.

\bibitem[{\citenamefont{Jentschura}(2011{\natexlab{b}})}]{Jentschura2011516}
\bibinfo{author}{\bibfnamefont{U.}~\bibnamefont{Jentschura}},
  \bibinfo{journal}{Annals of Physics} \textbf{\bibinfo{volume}{326}},
  \bibinfo{pages}{516 } (\bibinfo{year}{2011}{\natexlab{b}}), ISSN
  \bibinfo{issn}{0003-4916}, \eprint{1011.5453},
  \urlprefix\url{http://www.sciencedirect.com/science/article/pii/S00034916100%
02009}.

\bibitem[{\citenamefont{Dorokhov et~al.}(2020)\citenamefont{Dorokhov, Faustov,
  Martynenko, and Martynenko}}]{Dorokhov:2020ubu}
\bibinfo{author}{\bibfnamefont{A.}~\bibnamefont{Dorokhov}},
  \bibinfo{author}{\bibfnamefont{R.}~\bibnamefont{Faustov}},
  \bibinfo{author}{\bibfnamefont{A.}~\bibnamefont{Martynenko}},
  \bibnamefont{and}
  \bibinfo{author}{\bibfnamefont{F.}~\bibnamefont{Martynenko}}
  (\bibinfo{year}{2020}), \eprint{2010.07380}.

\bibitem[{\citenamefont{Matveev et~al.}(2013)\citenamefont{Matveev, Parthey,
  Predehl, Alnis, Beyer, Holzwarth, Udem, Wilken, Kolachevsky, Abgrall
  et~al.}}]{PhysRevLett.110.230801}
\bibinfo{author}{\bibfnamefont{A.}~\bibnamefont{Matveev}},
  \bibinfo{author}{\bibfnamefont{C.~G.} \bibnamefont{Parthey}},
  \bibinfo{author}{\bibfnamefont{K.}~\bibnamefont{Predehl}},
  \bibinfo{author}{\bibfnamefont{J.}~\bibnamefont{Alnis}},
  \bibinfo{author}{\bibfnamefont{A.}~\bibnamefont{Beyer}},
  \bibinfo{author}{\bibfnamefont{R.}~\bibnamefont{Holzwarth}},
  \bibinfo{author}{\bibfnamefont{T.}~\bibnamefont{Udem}},
  \bibinfo{author}{\bibfnamefont{T.}~\bibnamefont{Wilken}},
  \bibinfo{author}{\bibfnamefont{N.}~\bibnamefont{Kolachevsky}},
  \bibinfo{author}{\bibfnamefont{M.}~\bibnamefont{Abgrall}},
  \bibnamefont{et~al.}, \bibinfo{journal}{Phys. Rev. Lett.}
  \textbf{\bibinfo{volume}{110}}, \bibinfo{pages}{230801}
  (\bibinfo{year}{2013}),
  \urlprefix\url{https://link.aps.org/doi/10.1103/PhysRevLett.110.230801}.

\bibitem[{\citenamefont{Akhiezer and Berestetskii}(1965)}]{Akhiezer:1965engl}
\bibinfo{author}{\bibfnamefont{A.~I.} \bibnamefont{Akhiezer}} \bibnamefont{and}
  \bibinfo{author}{\bibfnamefont{V.~B.} \bibnamefont{Berestetskii}},
  \emph{\bibinfo{title}{{Quantum Electrodynamics}}}
  (\bibinfo{publisher}{Interscience Publishers}, \bibinfo{address}{New York},
  \bibinfo{year}{1965}).

\bibitem[{\citenamefont{Pilkuhn}(1979)}]{Pilkuhn1979}
\bibinfo{author}{\bibfnamefont{H.~M.} \bibnamefont{Pilkuhn}},
  \emph{\bibinfo{title}{Relativistic Particle Physics}}
  (\bibinfo{publisher}{Springer Berlin Heidelberg}, \bibinfo{address}{New
  York}, \bibinfo{year}{1979}), \bibinfo{edition}{1st} ed.

\bibitem[{\citenamefont{Lucha et~al.}(1991)\citenamefont{Lucha, Rupprecht, and
  Schoberl}}]{Lucha:1991jy}
\bibinfo{author}{\bibfnamefont{W.}~\bibnamefont{Lucha}},
  \bibinfo{author}{\bibfnamefont{H.}~\bibnamefont{Rupprecht}},
  \bibnamefont{and} \bibinfo{author}{\bibfnamefont{F.~F.}
  \bibnamefont{Schoberl}}, \bibinfo{journal}{Phys. Rev.}
  \textbf{\bibinfo{volume}{D44}}, \bibinfo{pages}{242} (\bibinfo{year}{1991}).

\bibitem[{\citenamefont{Galkin et~al.}(1992)\citenamefont{Galkin, Mishurov, and
  Faustov}}]{Galkin:1992ry}
\bibinfo{author}{\bibfnamefont{V.~O.} \bibnamefont{Galkin}},
  \bibinfo{author}{\bibfnamefont{A.~Y.} \bibnamefont{Mishurov}},
  \bibnamefont{and} \bibinfo{author}{\bibfnamefont{R.~N.}
  \bibnamefont{Faustov}}, \bibinfo{journal}{Sov. J. Nucl. Phys.}
  \textbf{\bibinfo{volume}{55}}, \bibinfo{pages}{1207} (\bibinfo{year}{1992}).

\bibitem[{\citenamefont{Crater et~al.}(1996)\citenamefont{Crater, Wong, and
  Wong}}]{Crater:1996ti}
\bibinfo{author}{\bibfnamefont{H.~W.} \bibnamefont{Crater}},
  \bibinfo{author}{\bibfnamefont{C.~W.} \bibnamefont{Wong}}, \bibnamefont{and}
  \bibinfo{author}{\bibfnamefont{C.-Y.} \bibnamefont{Wong}},
  \bibinfo{journal}{Int. J. Mod. Phys.} \textbf{\bibinfo{volume}{E5}},
  \bibinfo{pages}{589} (\bibinfo{year}{1996}), \eprint{hep-ph/9603402}.

\bibitem[{\citenamefont{Terekidi and Darewych}(2005)}]{Terekidi:2003gp}
\bibinfo{author}{\bibfnamefont{A.~G.} \bibnamefont{Terekidi}} \bibnamefont{and}
  \bibinfo{author}{\bibfnamefont{J.~W.} \bibnamefont{Darewych}},
  \bibinfo{journal}{J. Math. Phys.} \textbf{\bibinfo{volume}{46}},
  \bibinfo{pages}{032302} (\bibinfo{year}{2005}), \eprint{hep-ph/0311132}.

\bibitem[{\citenamefont{Bete and Salpeter}(1951)}]{Bete:1951}
\bibinfo{author}{\bibfnamefont{H.~A.} \bibnamefont{Bete}} \bibnamefont{and}
  \bibinfo{author}{\bibfnamefont{E.~E.} \bibnamefont{Salpeter}},
  \bibinfo{journal}{Phys. Rev.} \textbf{\bibinfo{volume}{84}},
  \bibinfo{pages}{1232} (\bibinfo{year}{1951}).

\bibitem[{\citenamefont{Salpeter}(1952)}]{Salpeter:1952}
\bibinfo{author}{\bibfnamefont{E.~E.} \bibnamefont{Salpeter}},
  \bibinfo{journal}{Phys. Rev.} \textbf{\bibinfo{volume}{87}},
  \bibinfo{pages}{328} (\bibinfo{year}{1952}).

\bibitem[{\citenamefont{Faustov et~al.}(1999)\citenamefont{Faustov,
  Karimkhodzhaev, and Martynenko}}]{Faustov:1998kx}
\bibinfo{author}{\bibfnamefont{R.~N.} \bibnamefont{Faustov}},
  \bibinfo{author}{\bibfnamefont{A.}~\bibnamefont{Karimkhodzhaev}},
  \bibnamefont{and} \bibinfo{author}{\bibfnamefont{A.~P.}
  \bibnamefont{Martynenko}}, \bibinfo{journal}{Phys. Atom. Nucl.}
  \textbf{\bibinfo{volume}{62}}, \bibinfo{pages}{2103} (\bibinfo{year}{1999}),
  \eprint{hep-ph/9808365}.

\bibitem[{\citenamefont{Faustov and Martynenko}(1998)}]{Faustov:1997rc}
\bibinfo{author}{\bibfnamefont{R.~N.} \bibnamefont{Faustov}} \bibnamefont{and}
  \bibinfo{author}{\bibfnamefont{A.~P.} \bibnamefont{Martynenko}},
  \bibinfo{journal}{Phys. Atom. Nucl.} \textbf{\bibinfo{volume}{61}},
  \bibinfo{pages}{471} (\bibinfo{year}{1998}), \eprint{hep-ph/9709374}.

\bibitem[{\citenamefont{Martynenko}(2006)}]{Martynenko:2006gz}
\bibinfo{author}{\bibfnamefont{A.~P.} \bibnamefont{Martynenko}}
  (\bibinfo{year}{2006}).

\bibitem[{\citenamefont{Polyzou et~al.}(2011)\citenamefont{Polyzou, Huang,
  Elster, Glockle, Golak, Skibinski, Witala, and Kamada}}]{Polyzou:2010kx}
\bibinfo{author}{\bibfnamefont{W.~N.} \bibnamefont{Polyzou}},
  \bibinfo{author}{\bibfnamefont{Y.}~\bibnamefont{Huang}},
  \bibinfo{author}{\bibfnamefont{C.}~\bibnamefont{Elster}},
  \bibinfo{author}{\bibfnamefont{W.}~\bibnamefont{Glockle}},
  \bibinfo{author}{\bibfnamefont{J.}~\bibnamefont{Golak}},
  \bibinfo{author}{\bibfnamefont{R.}~\bibnamefont{Skibinski}},
  \bibinfo{author}{\bibfnamefont{H.}~\bibnamefont{Witala}}, \bibnamefont{and}
  \bibinfo{author}{\bibfnamefont{H.}~\bibnamefont{Kamada}},
  \bibinfo{journal}{Few Body Syst.} \textbf{\bibinfo{volume}{49}},
  \bibinfo{pages}{129} (\bibinfo{year}{2011}), \eprint{1008.5215}.

\bibitem[{\citenamefont{Keister and Polyzou}(1991)}]{Keister:1991sb}
\bibinfo{author}{\bibfnamefont{B.~D.} \bibnamefont{Keister}} \bibnamefont{and}
  \bibinfo{author}{\bibfnamefont{W.~N.} \bibnamefont{Polyzou}},
  \bibinfo{journal}{Adv. Nucl. Phys.} \textbf{\bibinfo{volume}{20}},
  \bibinfo{pages}{225} (\bibinfo{year}{1991}).

\bibitem[{\citenamefont{Brown and Jackson}(1976)}]{Brown79engl}
\bibinfo{author}{\bibfnamefont{G.~E.} \bibnamefont{Brown}} \bibnamefont{and}
  \bibinfo{author}{\bibfnamefont{A.~D.} \bibnamefont{Jackson}},
  \emph{\bibinfo{title}{The Nucleon-nucleon interaction}}
  (\bibinfo{publisher}{North-Holland pubshing}, \bibinfo{address}{New York},
  \bibinfo{year}{1976}).

\bibitem[{\citenamefont{Klink}(2003)}]{Klink:2003ec}
\bibinfo{author}{\bibfnamefont{W.~H.} \bibnamefont{Klink}},
  \bibinfo{journal}{Few Body Syst.} \textbf{\bibinfo{volume}{33}},
  \bibinfo{pages}{99} (\bibinfo{year}{2003}).

\bibitem[{\citenamefont{Bjorken and Drell}(1964)}]{Bjorken64}
\bibinfo{author}{\bibfnamefont{J.~D.} \bibnamefont{Bjorken}} \bibnamefont{and}
  \bibinfo{author}{\bibfnamefont{S.}~\bibnamefont{Drell}},
  \emph{\bibinfo{title}{Relativistic Quantum Mechanics}},
  vol.~\bibinfo{volume}{1} (\bibinfo{publisher}{McGraw-Hill.},
  \bibinfo{address}{Berlin-G\"{o}ttingen-Heidelberg}, \bibinfo{year}{1964}).

\bibitem[{\citenamefont{Bodwin and Yennie}(1988)}]{Bodwin:1987mj}
\bibinfo{author}{\bibfnamefont{G.~T.} \bibnamefont{Bodwin}} \bibnamefont{and}
  \bibinfo{author}{\bibfnamefont{D.~R.} \bibnamefont{Yennie}},
  \bibinfo{journal}{Phys. Rev.} \textbf{\bibinfo{volume}{D37}},
  \bibinfo{pages}{498} (\bibinfo{year}{1988}).

\bibitem[{\citenamefont{Andreev}(2003)}]{Andreev2003ce}
\bibinfo{author}{\bibfnamefont{V.~V.} \bibnamefont{Andreev}},
  \bibinfo{journal}{Physics of Atomic Nuclei} \textbf{\bibinfo{volume}{66}},
  \bibinfo{pages}{383} (\bibinfo{year}{2003}), ISSN \bibinfo{issn}{1063-7788},
  \bibinfo{note}{10.1134/1.1553511},
  \urlprefix\url{http://dx.doi.org/10.1134/1.1553511}.

\bibitem[{\citenamefont{Andreev}(2004)}]{Andreev:2004bn}
\bibinfo{author}{\bibfnamefont{V.}~\bibnamefont{Andreev}}, in
  \emph{\bibinfo{booktitle}{{18th International Workshop on High-Energy Physics
  and Quantum Field Theory}}} (\bibinfo{year}{2004}), pp.
  \bibinfo{pages}{148--153}, \eprint{hep-ph/0407055}.

\bibitem[{\citenamefont{Bellomo}(1961)}]{Bellomo61}
\bibinfo{author}{\bibfnamefont{E.}~\bibnamefont{Bellomo}}, \bibinfo{journal}{Il
  Nuovo Cimento} \textbf{\bibinfo{volume}{21}}, \bibinfo{pages}{730}
  (\bibinfo{year}{1961}).

\bibitem[{\citenamefont{Bogush and Fedorov}(1962)}]{Bogush1962a}
\bibinfo{author}{\bibfnamefont{A.~A.} \bibnamefont{Bogush}} \bibnamefont{and}
  \bibinfo{author}{\bibfnamefont{F.~I.} \bibnamefont{Fedorov}},
  \bibinfo{journal}{Vesti AN BSSR} \textbf{\bibinfo{volume}{ser.fiz.-m.n.}},
  \bibinfo{pages}{26} (\bibinfo{year}{1962}), \bibinfo{note}{in Russian}.

\bibitem[{\citenamefont{Gastmans and Wu}(1990)}]{Gastmans:1990xh}
\bibinfo{author}{\bibfnamefont{R.}~\bibnamefont{Gastmans}} \bibnamefont{and}
  \bibinfo{author}{\bibfnamefont{T.~T.} \bibnamefont{Wu}},
  \emph{\bibinfo{title}{The Ubiquitous photon: Helicity method for QED and
  QCD}} (\bibinfo{publisher}{Oxford, UK}, \bibinfo{year}{1990}).

\bibitem[{\citenamefont{Dittmaier}(1999)}]{Dittmaier:1998nn}
\bibinfo{author}{\bibfnamefont{S.}~\bibnamefont{Dittmaier}},
  \bibinfo{journal}{Phys. Rev.} \textbf{\bibinfo{volume}{D59}},
  \bibinfo{pages}{016007} (\bibinfo{year}{1999}), \eprint{hep-ph/9805445}.

\bibitem[{\citenamefont{Andreev}(2009)}]{Andreev2009z}
\bibinfo{author}{\bibfnamefont{V.~V.} \bibnamefont{Andreev}},
  \bibinfo{journal}{Nonlinear phenomena in complex systems}
  \textbf{\bibinfo{volume}{12}}, \bibinfo{pages}{338} (\bibinfo{year}{2009}).

\end{thebibliography}
\end{document}